\documentclass[aps,prc,preprint,groupedaddress]{revtex4-1}

\usepackage[]{graphicx}	
\usepackage{braket}
\usepackage{xspace}
\usepackage{CJK}
\usepackage{epstopdf}

\newcommand{\dAP}{d_{\alpha\mbox{-}{}^{31}\mathrm{P}}}
\newcommand{\dtS}{d_{t\mbox{-}{}^{32}\mathrm{S}}}
\newcommand{\dboth}{d_{\alpha\mbox{-}{}^{31}\mathrm{P},t\mbox{-}{}^{32}\mathrm{S}}}
\newcommand{\AP}{$\alpha$-$^{31}$P\xspace}
\newcommand{\tS}{$t$-$^{32}$S\xspace}

\begin{document}


\title{$\alpha$ and triton clustering in $^{35}$Cl}



 \author{Yasutaka Taniguchi}
\affiliation{Department of Information Engineering, National Institute of Technology, Kagawa College, Kagawa 769-1192, Japan}


\date{\today}

\begin{abstract}
\begin{description}
 \item[Background] 
	    Coupling of cluster and deformed structures are important for dynamics of nuclear structure.
	    Threshold energy has been discussed to explain cluster structures  coupling to deformed states but relation between threshold energy and excitation energy has open problems.
	    Negative-parity superdeformed (SD) states were observed by a $\gamma$-spectroscopy experiment in $^{35}$Cl but its detailed structure is unclear.
 \item[Purpose]
	    By analyzing coupling of cluster structures in deformed states and high-lying cluster states in $^{35}$Cl, cluster structures coupling to deformed states and excitation energy of high-lying cluster states are investigated.
 \item[Method]
	    The antisymmetrized molecular dynamics (AMD) and the generator coordinate method (GCM) are used.
	    An AMD wave function is a Slater determinant of Gaussian wave packets.
	    By energy variational calculations with constraints on deformation and clustering, wave functions of deformed structures and \AP and \tS cluster structures are obtained.
	    Adopting those wave functions as GCM basis, wave functions of ground and excited states are calculated.
 \item[Results]
	    Various deformed bands are obtained and predicted.
	    A $K^\pi = \frac{1}{2}^-$ deformed band, which corresponds to the observed SD band, dominates deformed structure and compact \AP and \tS cluster structure components.
	    Particle-hole configurations of the dominant components with deformed and cluster structures are similar.
	    In high-lying states, almost pure \AP and \tS cluster states are obtained in negative-parity states, and excitation energies of the \tS cluster states are higher than those of \AP cluster states.
 \item[Conclusions]
	    Particle-hole configurations of cluster structure with small intercluster distance are important for coupling to low-energy deformed states.
	    Threshold energies reflect to excitation energies of high-lying almost pure cluster states.
\end{description}
\end{abstract}


\maketitle

\section{Introduction}

Nuclear structure changes drastically by excitation.
A goal of nuclear physics is to understand mechanism of structure changes.
Clustering and deformation play important roles for nuclear structure.
For examples, $\alpha$-clustering explains coexistence of inversion doublet bands\cite{10.1143/PTP.40.277} and enhancement of $\alpha$-transfer\cite{BECCHETTI1978313,ANANTARAMAN1979445,PhysRevC.93.034606,FUKUI201938} and -knockout reaction cross sections\cite{PhysRevC.15.69,PhysRevC.22.1394,PhysRevC.31.1662,PhysRevC.40.1130,PhysRevC.98.024614}. 
Deformation explains strong E2 transition strengths and rotational spectra.

Cluster structures couple to deformed states in $sd$- and $pf$-shell region.
In $^{40}$Ca, normal-deformed (ND) and superdeformed (SD) bands have been observed up to high-spin states\cite{PhysRevLett.87.222501}.
$\alpha$-$^{36}$Ar clustering of the ND band was predicted by cluster model\cite{PTP.80.598,PhysRevC.41.63}, and it was confirmed experimentally by $\alpha$-transfer reaction\cite{Yamaya1994154}.
Coupling of $\alpha$-$^{36}$Ar cluster structure to the ND states were discussed by using semi-microscopic\cite{PhysRevC.49.149} and full microscopic calculations\cite{taniguchi:044317}.
Coupling and mixing of cluster structures to deformed states are also discussed in $^{32,34}$S\cite{PhysRevC.66.021301,PhysRevC.69.051304}, $^{36}$Ar\cite{Sakuda200477}, $^{42}$Ca\cite{PhysRevC.51.586,Taniguchi01072014} and $^{44}$Ti\cite{FRIEDRICH197547,PhysRevC.41.63,PhysRevC.42.1935,Kimura200658}.

In order to understand cluster structure, the threshold rule has been proposed\cite{doi:10.1143/PTPS.E68.464}.
The threshold rule predicts that cluster structures are developed in excited states whose excitation energies are similar to threshold energies of cluster decay.
It is powerful in very light nuclei such as $^{12}$C and $^{16}$O.
The $J^\pi = 0_2^+$ state in $^{12}$C is considered to have dilute $3\alpha$ structure\cite{10.1143/PTP.57.1262,10.1143/PTP.62.1621,KAMIMURA1981456,PhysRevC.67.051306}, and its excitation energy (7.65~MeV) is similar to $3\alpha$ threshold energy (7.27~MeV).
In $^{16}$O, $\alpha$-$^{12}$C cluster structure develops at the $J^\pi = 0_2^+$ state (6.04~MeV)\cite{10.1143/PTP.40.277,10.1143/PTP.51.1621,10.1143/PTP.55.1751,10.1143/PTP.56.111}, which is close to $\alpha + {}^{12}\textrm{C}$ threshold energy (7.16~MeV).
In $sd$-shell or heavier region, however, the threshold rule do not work quantitatively.
The $J^\pi = 0_2^+$ state in $^{40}$Ca is considered to contain a large amount of $\alpha$-$^{36}$Ar cluster structure components, but its excitation energy (3.35~MeV) is much less than $\alpha$ + $^{36}$Ar threshold energy (7.04~MeV).

Particle-hole excitations are also discussed for nuclear clustering.
When nucleons are excited to higher shell and those nucleons correlate spatially and strongly, they are expected to form a cluster\cite{ARIMA1967129,doi:10.1143/PTPS.52.1}.
However, relations of intercluster motion and particle-hole configurations are open problems.

The threshold rule is simple, and it has been widely discussed in clustering in deformed states.
In $^{35}$Cl, a negative-parity superdeformed band have been observed from $J^\pi = \frac{15}{2}^-$ (8.31~MeV) to $\frac{27}{2}^{(-)}$ (16.30~MeV) states by a $\gamma$-spectroscopy experiment\cite{PhysRevC.88.034303}.
In-band $B(\mathrm{E2})$ values are deduced, and the values are around 30~W.u, which shows the band form largely deformed structure.
In Ref.~\onlinecite{PhysRevC.88.034303}, $\alpha$-$^{32}$S cluster structure with one proton hole of the band is discussed with mentioning threshold energies of $\alpha + {}^{31}$P and $t + {}^{32}$S channels, which are 6.99~MeV and 17.94~MeV, respectively.
Just because of lower threshold energy, coupling of \AP cluster structure to the SD band is predicted. 
However, details of cluster coupling of the band are not discussed, and theoretical work of cluster structure are insufficient in $^{35}$Cl.

Naive structure of deformed states are described by using the Nilsson model, which shows single-particle orbits on a deformed mean-field.
By analyzing single-particle orbits on a deformed mean-field, multi-particle multi-hole configurations of largely deformed states are explained such as SD states in $^{40}$Ca\cite{PhysRevLett.87.222501,Inakura2002261,PhysRevC.68.044321}, but it negletcs many-body correlation effects such as clustering.
In order to investigate detailed structure including correlation effects, multi-particle correlation effects should be taken into account directly.

This paper aims to clarify $\alpha$-$^{31}$P and $t$-$^{32}$S cluster correlations in $^{35}$Cl.
By using the antisymmetrized molecular dynamics (AMD) and the generator coordinate method (GCM), various structures including SD states are obtained.
Cluster structures in the SD band and high-lying states are analyzed focusing on particle-hole configurations and threshold energies, and coupling of cluster structure to deformed states and mechanism to generate high-lying cluster states are investigated.

This paper is organized as follows.
In Sec.~\ref{framework}, the framework of this works is explained briefly.
In Sec.~\ref{results}, numerical results about energies of deformed and cluster structure, level scheme, $B(\mathrm{E2})$ values, and the amounts of \AP and \tS cluster structure components are shown.
In Sec.~\ref{discussions}, cluster structures coupling to deformed states and excitation energies of high-lying cluster states are discussed.
Finally, conclusions are given in Sec.~\ref{conclusions}.


\section{Framework}
\label{framework}

In this section, the framework of the study is explained briefly.
The details of the framework are provided in Refs.~\onlinecite{PTP.93.115,PhysRevC.69.044319,PTP.112.475}.

\subsection{Wave function}

The wave functions are obtained by the GCM after parity and angular momentum projection using deformed-basis AMD wave functions.
 A deformed-basis AMD wave function $\ket{\mathrm{\Phi}}$ is a Slater determinant of Gaussian wave packets that can deform triaxially such that
\begin{eqnarray}
 &\ket{\mathrm{\Phi}} = \hat{\cal A}\ket{\varphi_1, \varphi_2, \cdots, \varphi_A},&\\
 &\ket{\varphi_i} = \ket{\phi_i} \otimes \ket{\chi_i} \otimes \ket{\tau_i},& \\
 &\langle \mathbf{r} | \phi_i \rangle = \left(\frac{\det \mathsf M}{\pi^3}\right)^{1/4} \exp\left[ - \frac{1}{2} (\mathbf{r} - \mathbf{Z}_i) \cdot \mathsf{M} (\mathbf{r} - \mathbf{Z}_i)\right],& \\
 &\ket{\chi_i} = \chi^\uparrow_i \ket{\uparrow} + \chi^\downarrow_i \ket{\downarrow},& \\
 &\ket{\tau_i} = \ket{\pi}\ \mathrm{or}\ \ket{\nu},&
\end{eqnarray}
where $\hat{\cal A}$ denotes the antisymmetrization operator, and
$\ket{\varphi_i}$ denotes a single-particle wave function. Further,
$\ket{\phi_i}$, $\ket{\chi_i}$, and $\ket{\tau_i}$ denote the
spatial, spin, and isospin components, respectively, of each
single-particle wave function $\ket{\varphi_i}$. The real $3 \times
3$ matrix $\mathsf{M}$ denotes the width of the Gaussian
single-particle wave functions that can deform triaxially and is
common to all nucleons. $\mathbf{Z}_i = (Z_{ix}, Z_{iy}, Z_{iz})$ are complex parameters denoting the centroid of
each single-particle wave function in phase space. The complex
parameters $\chi_i^\uparrow$ and $\chi_i^\downarrow$ denote spin direction.
The isospin component of each single-particle wave function is fixed as a proton ($\pi$) or a neutron ($\nu$).
 Axial symmetry is not assumed.

\subsection{Energy variation}

The intrinsic wave functions of the GCM basis are obtained by energy
variation with a constraint potential $V_\mathrm{cnst}$ after projection onto eigen states of parity ($\pi = \pm 1$),
\begin{eqnarray}
 &\delta \left( \frac{\braket{\Phi^\pi | \hat{H} | \Phi^\pi}}{\braket{\Phi^\pi | \Phi^\pi}} + V_{\mathrm{cnst}} \right) = 0,&\\
 &\ket{\Phi^\pi} = \frac{1 + \pi \hat{P}_r}{2} \ket{\Phi},&
\end{eqnarray}
where $\hat{H}$ and $\hat{P}_r$ denote Hamiltonian and parity operators, respectively.
The variational parameters are
$\mathsf{M}$, $\mathbf{Z}_i$, and $\chi_i^{\uparrow,\downarrow}$ ($i= 1, ..., A$).
 The Gogny D1S force is used as an effective interaction. 
Variational calculations are performed by using the conjugate gradient method.

In order to obtain the various wave functions, constraint potentials $V_{\mathrm{cnst}}$ with parabola form are added.
In this work, two kinds of constraint potentials are used, which are for the matter quadrupole deformation parameter $\beta$ of the total system ($\beta$-constraint) and intercluster distance.
For intercluster distance, distance between $\alpha$ and $^{31}$P ($\dAP$) or between $t$ and $^{32}$S ($\dtS$) is constrained.

  \subsection{Generator coordinate method}

  By using the GCM, optimized wave functions are superposed after parity and angular momentum projection,
 \begin{equation}
    \Ket{\Phi^{J\pi}_M} = \sum_i \hat{P}_{MK_i}^{J\pi}  \Ket{\Phi^{c_i}(q_i)} f_i , \label{eq:HW}
 \end{equation} 
where $\hat{P}_{MK}^{J^\pi}$ is the parity and total angular momentum
  projection operator, and $\Ket{\Phi^{c_i}(q_i)}$ is a basis wave function.
  The $c_i$ is a kind of constraint potential ($\beta$, $\dAP$, and $\dtS$), and it is constrained to $q_i$ in the variational calculation.
  The integrals over the three Euler angles in the angular momentum projection operator $\hat{P}_{MK}^J$ are evaluated by numerical integration.  
  The numbers of sampling points $(N_\alpha, N_\beta, N_\gamma)$ in the numerical integration for Euler angles $\alpha$, $\beta$, and $\gamma$, respectively, are $(N_\alpha, N_\beta, N_\gamma) = (29, 25, 29)$ and $(26, 25, 26)$ for positive- and negative-parity states, respectively. 
  Here the body-fixed $z$-axis is determined by minimizing variance of $\hat{J}_z$, which is $z$ component of angular momentum\cite{Taniguchi01102016}.
  The coefficients $f_i$ are determined by the Hill--Wheeler equation,  
  \begin{equation}
    \delta \left( \Braket{\Phi^{J\pi}_M | \hat{H} | \Phi^{J\pi}_M} - \epsilon \Braket{\Phi^{J\pi}_M | \Phi^{J\pi}_M}\right) = 0. 
  \end{equation}

\section{Results}
\label{results}

\begin{figure}[tbp]
 \includegraphics[width=0.45\textwidth]{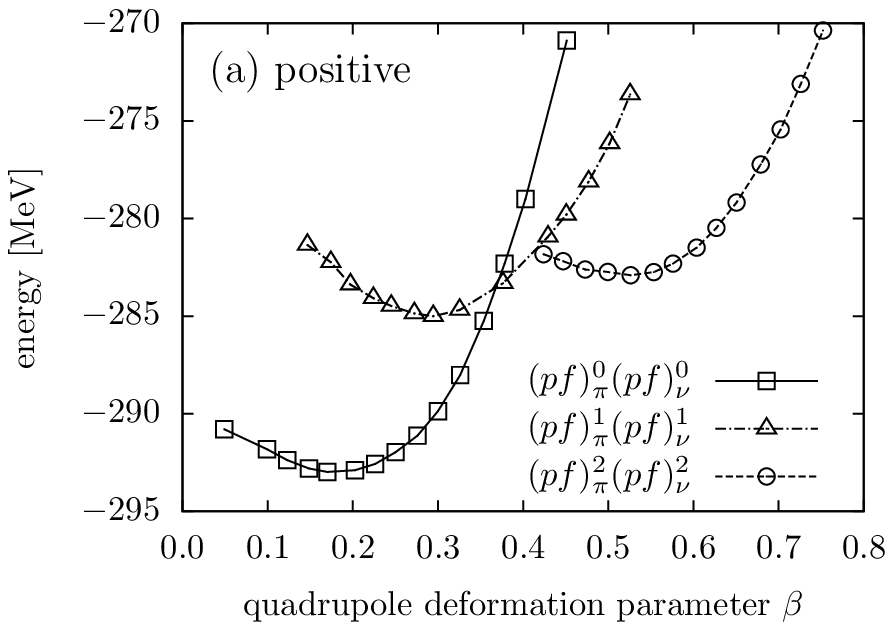}\\
 \includegraphics[width=0.45\textwidth]{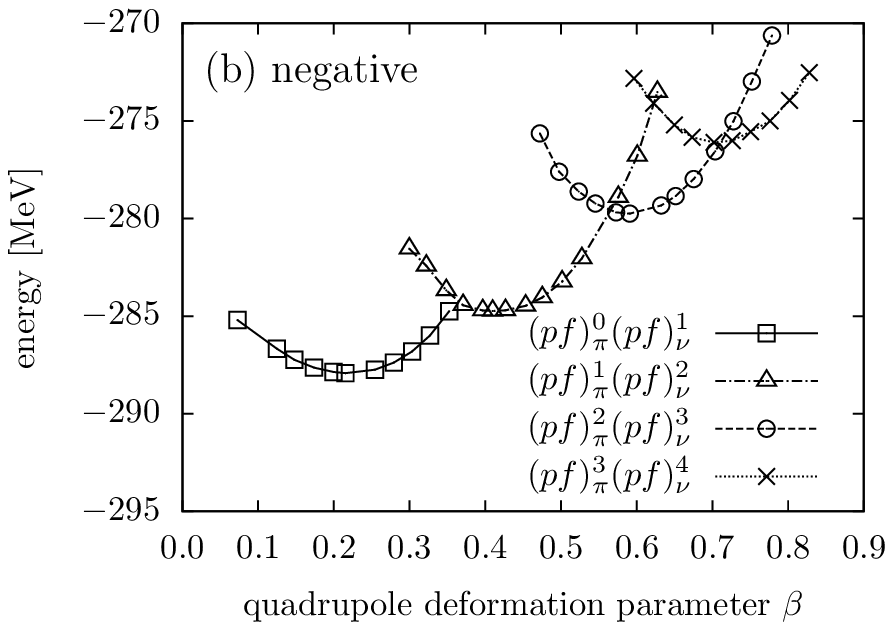}
\caption{
 Energy curves as functions of the quadrupole deformation parameter $\beta$ obtained by energy variational calculation with parity projection for (a) positive- and (b) negative-parity.
 $(pf)^n_\tau$ ($\tau = \pi$ or $\nu$) shows particle-hole configurations (See text).
 For positive-parity states, squares, triangles, and circles are for totally $0\hbar\omega$, $2\hbar\omega$, and $4\hbar\omega$ excited configurations, respectively, from the lowest allowed state.
 For negative-parity states, squares, triangles, circles, and crosses are for totally $1\hbar\omega$, $3\hbar\omega$, $5\hbar\omega$, and $7\hbar\omega$ excited configurations, respectively, from the lowest allowed state.
 }
\label{beta_energy}
\end{figure}

Figure~\ref{beta_energy} shows energy curves as functions of the quadrupole deformation parameter $\beta$ obtained by energy variational calculation with the $\beta$-constraint after parity projection.
In both parity states, various structures are obtained. 
$(pf)^n_\tau$ ($\tau = \pi$ or $\nu$) denotes a particle-hole configuration, which shows that $n$ protons ($\pi$) or neutrons ($\nu$) are excited from $sd$-shell originated Nilsson orbits to the $pf$-shell originated ones.
All obtained wave functions with the $\beta$-constraint have no necked structures, which are called ``deformed structures'' following in this paper.
In positive-parity state [Fig.~\ref{beta_energy}(a)], the lowest energy state exist at $\beta \simeq 0.2$ in which a proton nor a neutron excite to the $pf$-shell.
In largely deformed region, wave functions that have $(pf)_\pi^1 (pf)_\nu^1$ and $(pf)_\pi^2 (pf)_\nu^2$ configurations appear, which are totally $2\hbar\omega$ and $4\hbar\omega$ excited configurations, respectively.
The local minima of $2\hbar\omega$ and $4\hbar\omega$ excited deformed states are at $\beta \simeq 0.3$ and 0.5, respectively.
In negative-parity states [Fig.~\ref{beta_energy}(b)], four configurations with $(pf)_\pi^0 (pf)^1_\nu$, $(pf)_\pi^1 (pf)_\pi^2$, $(pf)_\pi^2 (pf)_\pi^3$, and $(pf)_\pi^3 (pf)_\pi^4$ are obtained on the $\beta$-energy surface, which are totally $1\hbar\omega$, $3\hbar\omega$, $5\hbar\omega$, and $7\hbar\omega$ excited configurations, respectively.
The $1\hbar\omega$, $3\hbar\omega$, $5\hbar\omega$, and $7\hbar\omega$ excited states have local minima at $\beta \simeq 0.2$, 0.4, 0.6, and 0.7, respectively.

\begin{figure*}[tbp]
 \begin{tabular}{cc}
  \includegraphics[width=0.475\textwidth]{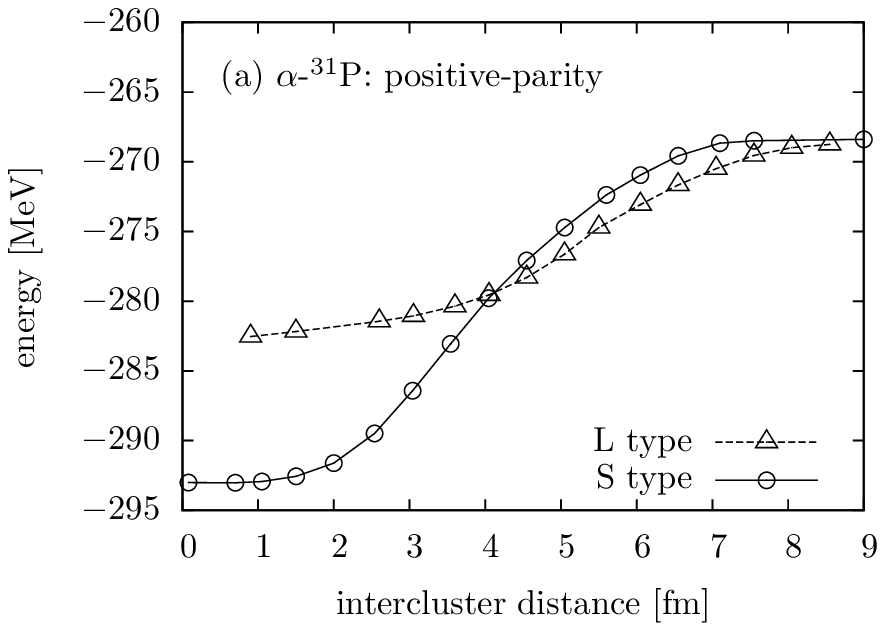}&
  \includegraphics[width=0.475\textwidth]{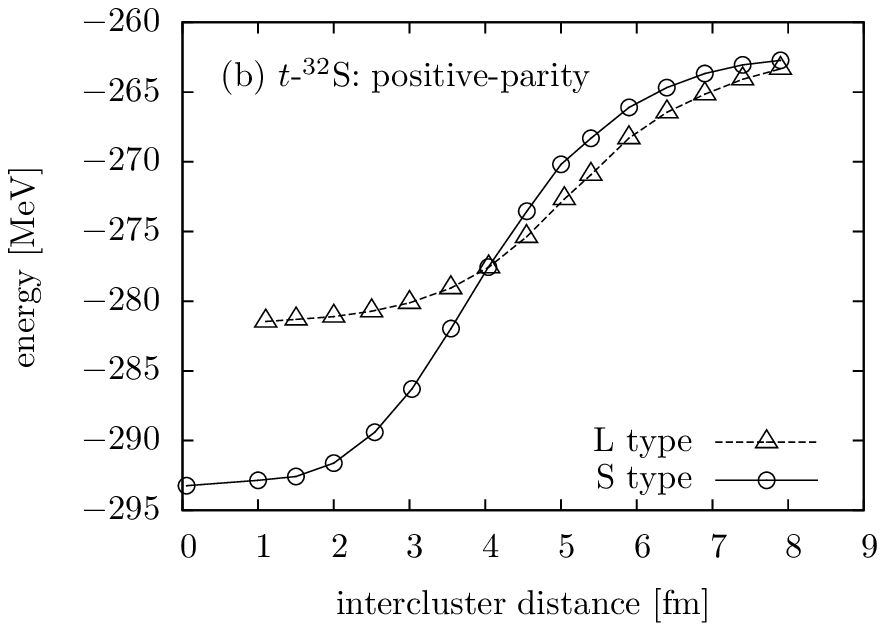}\\
  \includegraphics[width=0.475\textwidth]{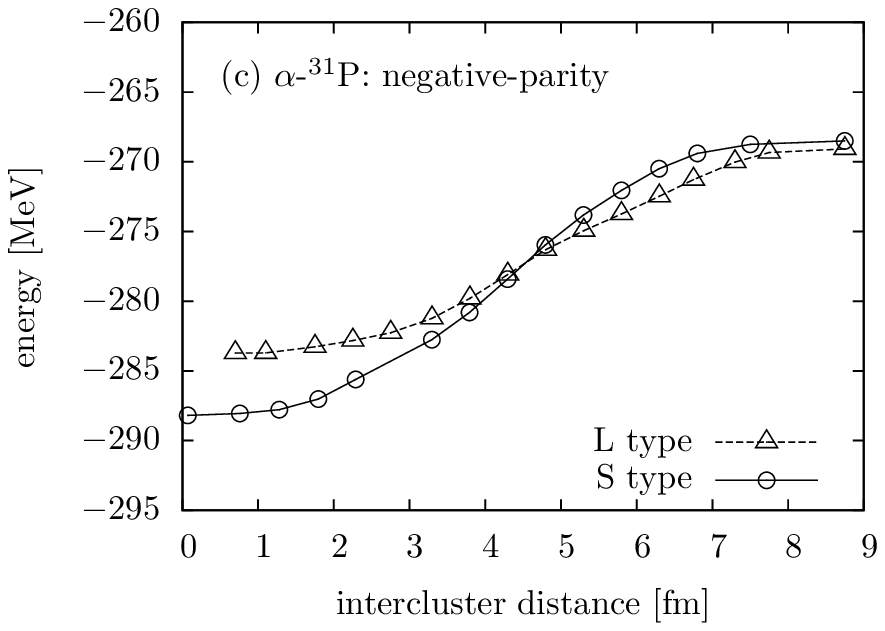}&
  \includegraphics[width=0.475\textwidth]{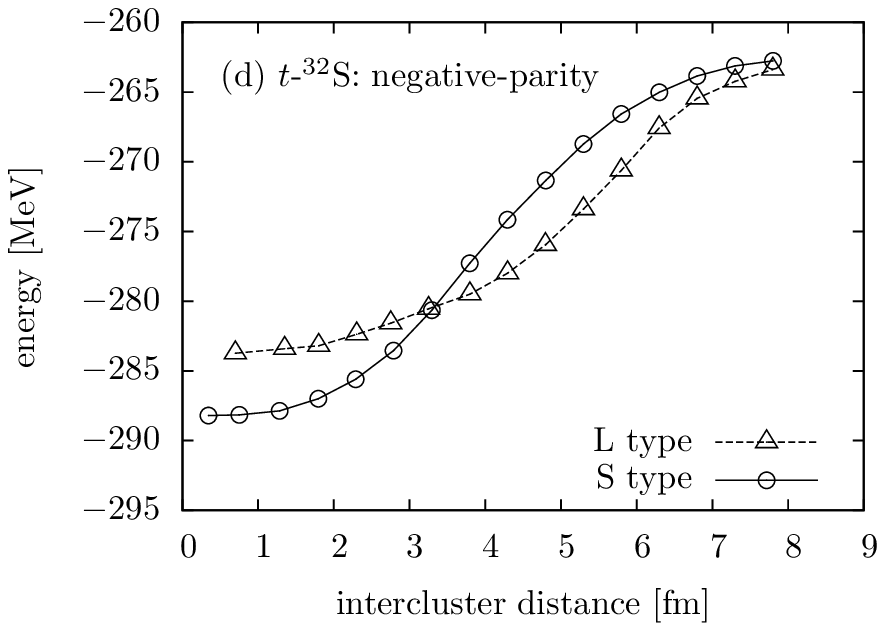}\\
 \end{tabular}
 \caption{
 Energies of cluster structure as functions of intercluster distance for positive-parity (a) \AP and (b) \tS and negative-parity (c) \AP and (d) \tS structures.
 Circles and triangles show energies of S- and L-types, respectively (see text).
 }
 \label{alpha}
\end{figure*}

\begin{figure}[tbp]
\begin{tabular}{cc}
 \includegraphics[width=0.2\textwidth]{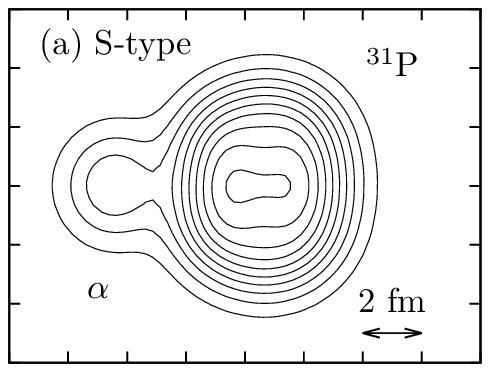}&
 \includegraphics[width=0.2\textwidth]{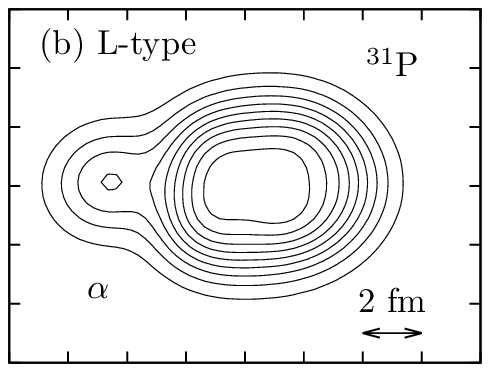}\\
\end{tabular} 
\caption{
 Density distributions of \AP cluster states in negative-parity states with $\dAP = 4.8$~fm for (a) S- and (b) L-types.
 Left and right peaks correspond to $\alpha$ and $^{31}$P clusters, respectively, in each panel.
 }
 \label{density}
\end{figure}

Figure~\ref{alpha} shows energies of $\alpha$-$^{31}$P and $t$-$^{32}$S cluster structures as functions of intercluster distance $\dAP$ and $\dtS$, respectively, which are obtained by energy variational calculations with constraints on $\dAP$ and $\dtS$, respectively.
By the energy variational calculations, two types of structures are obtained labeled S- and L-types.
The difference of the types is in orientation of deformed larger clusters, $^{31}$P and $^{32}$S.
In S- and L-type wave functions, smaller clusters, $\alpha$ and $t$, locate on the short and long axes of deformed larger clusters, respectively.
For example, a $^{31}$P cluster is deformed, and an $\alpha$ cluster is located on the short and long axes for S- and L-types as shown in Figs.~\ref{density}(a) and (b), respectively.
In Fig.~\ref{density}(a) and (b), long axes of $^{31}$P are direction of vertical and horizontal axes, respectively, and $\alpha$ clusters locate on the left side of the $^{31}$P clusters.
It shows that the $\alpha$ clusters locate on the short and long axes of $^{31}$P clusters for S- and L-types, respectively.
In short distance region, energies of same type configurations are similar for $\alpha$-$^{31}$P and $t$-$^{32}$S cluster structures as shown in Fig.~\ref{alpha}.
In positive-parity states, energies of S- and L-type structures with short intercluster distance are around $-293$ and $-283$~MeV, respectively, which are similar to energies of minimum energies of $0\hbar\omega$ and $4\hbar\omega$ excited configurations, respectively, on the $\beta$-energy surface [Fig.~\ref{beta_energy}(a)].
In negative-parity states, energies of S- and L-type structure with short intercluster distance are around $-288$ and $-284$~MeV, respectively, for both of $\alpha$-$^{31}$P and $t$-$^{32}$S cluster structures, which are similar to those of minimum energies of $1\hbar\omega$ and $3\hbar\omega$ excited configurations, respectively, on the $\beta$-energy surface [Fig.~\ref{beta_energy}(b)].
In large distance region, energies S- and L-type wave functions of same cluster structures are similar, for both parities.
Energies of \AP and \tS are $-269$ and $-263$ MeV, respectively, around Coulomb barrier, in which energies of $\alpha$-$^{31}$P cluster structures are lower than those of $t$-$^{32}$S cluster structure.
It reflects that threshold energy of $\alpha + {}^{31}$P is lower than that of $t+{}^{32}$S.
When intercluster distance is long, excitation energy of cluster structure is determined by threshold energy and Coulomb energy between clusters.

\begin{figure*}[tbp]
\begin{tabular}{cc}
 \includegraphics[width=.475\textwidth]{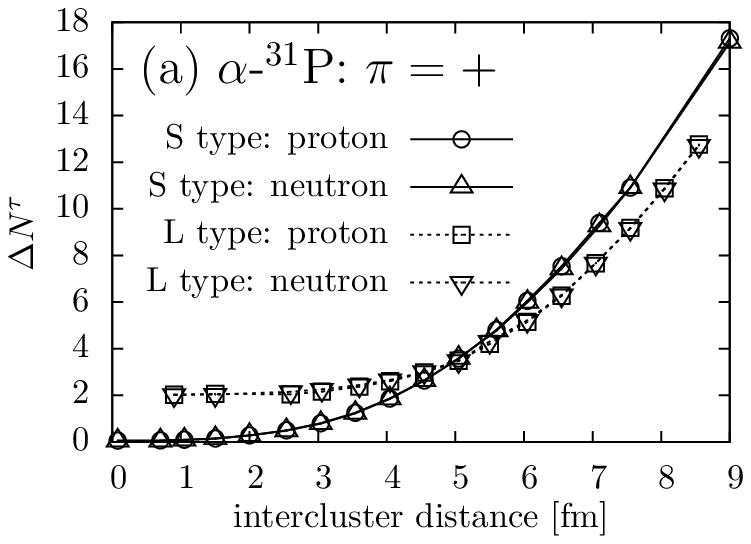}&
 \includegraphics[width=.475\textwidth]{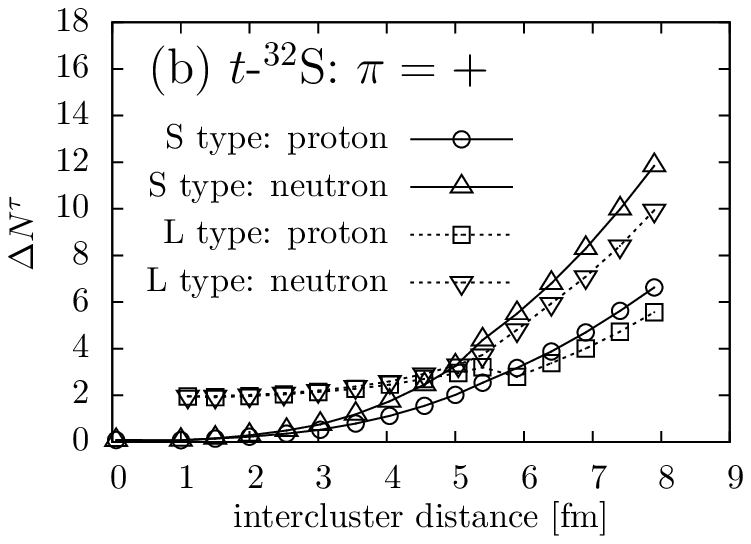}\\
 \includegraphics[width=.475\textwidth]{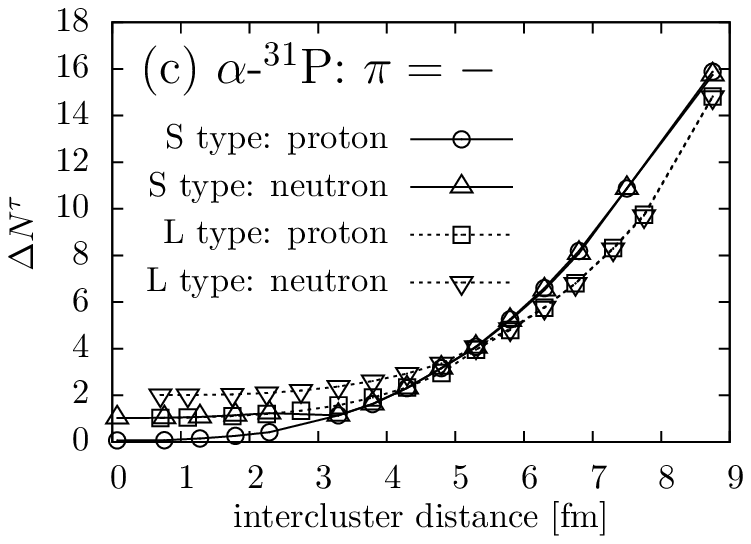}&
 \includegraphics[width=.475\textwidth]{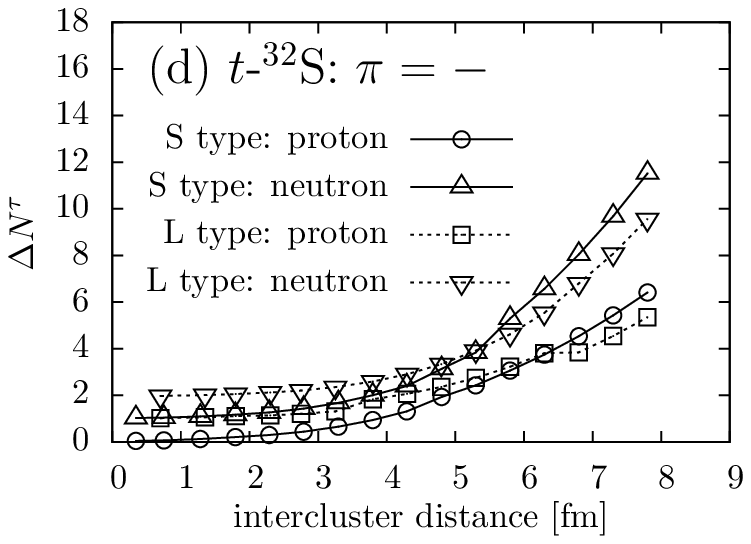}\\
\end{tabular} 
\caption{
 Relative harmonic oscillator quanta $\Delta N^\tau$ of positive-parity states of (a) $\alpha$-$^{31}$P and (b) $t$-$^{32}$S cluster structure and negative-parity states of (c) $\alpha$-$^{31}$P and (d) $t$-$^{32}$S cluster structure as functions of intercluster distance.
 Circles and triangles are for S-type protons and neutrons, respectively.
 Squares and inverted triangles are for L-type protons and neutrons, respectively.
 }
 \label{ho}
\end{figure*}

Figure~\ref{ho} shows relative deformed harmonic oscillator quanta $\Delta N^\tau$ of $\alpha$-$^{31}$P and $t$-$^{32}$S cluster structures as functions of intercluster distance, which are defined as
\begin{equation}
 \Delta N^\tau = \Braket{\sum_{i \in \tau}
  \left(
  \frac{1}{2} \hat{\mathbf{k}}_i \cdot \mathsf{M}^{-1} \hat{\mathbf{k}}_i + \frac{1}{2} \hat{\mathbf{r}}_i \cdot \mathsf{M} \hat{\mathbf{r}}_i - \frac{3}{2}
  \right)
  } - N_0^\tau.
\end{equation}
$\tau$ is $\pi$ (proton) or $\nu$ (neutron), and $i \in \tau$ shows that expectation values are summed up for protons or neutrons.
$\hat{\mathbf{r}}_i$ and $\hat{\mathbf{k}}_i$ are coordinate and wave number operators, respectively.
$N_0^\tau$ denotes harmonic oscillator quanta of the lowest allowed states of $^{35}$Cl, which are 24 and 26 for protons and neutrons, respectively.

In small intercluster distance region, $\Delta N^\tau$ values of \AP and \tS are similar for both parity and types.
Positive-parity S- and L-type wave functions have $(\Delta N^\pi, \Delta N^\nu) = (0, 0)$ and $(2, 2)$ configurations, respectively, and negative-parity S- and L-type wave functions have $(\Delta N^\pi, \Delta N^\nu) = (0, 1)$ and $(1, 2)$ configurations, respectively.
In details, the $(\Delta N^\pi, \Delta N^\nu) = (0, 0)$, $(2, 2)$, $(0, 1)$ and $(1, 2)$ configurations have the $(pf)_\pi^0(pf)_\nu^0$, $(pf)_\pi^2(pf)_\nu^2$, $(pf)_\pi^0(pf)_\nu^1$, and $(pf)_\pi^1(pf)_\nu^2$ configurations, respectively.
Energies of cluster structures with short intercluster distance region are similar to local minimum energies on the $\beta$-energy curves (Fig.~\ref{beta_energy}) for each corresponding particle-hole configuration.

Except for a proton part of positive-parity L-type \tS states, $\Delta N^\tau$ values as functions of intercluster distance $\dAP$ and $\dtS$ increase smoothly.
It shows that internal wave functions of clusters do not change drastically.
By superposition of wave functions with various intercluster distance, intercluster motion is taken into account.
In cluster structures with short intercluster distance, particle-hole configurations are determined by deformation of clusters and antisymmetrization effects.
For example, in the case of a negative-parity L-type \tS cluster structure, a $^{32}$S cluster is deformed and a long axis of the $^{32}$S is fully occupied up to the $sd$-shell.
When a $t$ cluster approaches to the $^{32}$S cluster on the long axis, three nucleons of the $t$ cluster cannot go into the $sd$-shell due to antisymmetrization effects and are left to the $pf$-shell.
It has a $3\hbar\omega$ excited configuration.
In the case of a negative-parity L-type \AP cluster structure, structure of $^{31}$P is almost the same as $^{32}$S but $^{31}$P cluster has one proton hole on the direction of long axis.
When an $\alpha$ cluster approach to the $^{31}$P cluster on the long axis, one proton in the $\alpha$ cluster can go into the proton hole of the $sd$-shell and other three nucleons in the $t$ cluster are left to the $pf$-shell.
It has also a $3\hbar\omega$ excited configuration.

In contrast, $\Delta N^\pi$ values of positive-parity L-type \tS states decrease at $\dtS\sim 5.5~\mathrm{fm}$ and increase again for $\dtS \gtrsim 6.0~\mathrm{fm}$, which shows that internal wave functions of clusters change drastically at $\dtS\sim 5.5$--6.0~fm.
In $\dtS \gtrsim 6.0~\mathrm{fm}$, internal wave functions of clusters are similar to a Hartree-Fock state due to weak interaction between clusters.
In $\dtS \lesssim 5.5~\mathrm{fm}$, $t$ and $^{32}$S clusters are distorted and $\Delta N^\pi$ become larger value.
It is because undistorted L-type \tS cluster structures with small intercluster distance can contain only negative-parity components.
The $sd$-shell of an undistorted $^{32}$S cluster is fully occupied up to the $sd$-shell in direction of the long axis, so three nucleons of $t$ cluster should go into the $pf$-shell in an undistorted \tS cluster structure with small intercluster distance.
As a result, total system of L-type undistorted \tS cluster structure with small intercluster cluster distance have $(pf)_\pi^1 (pf)_\nu^2$ configurations, which are negative-parity states.


\begin{figure*}[tbp]
 \includegraphics[width=\textwidth]{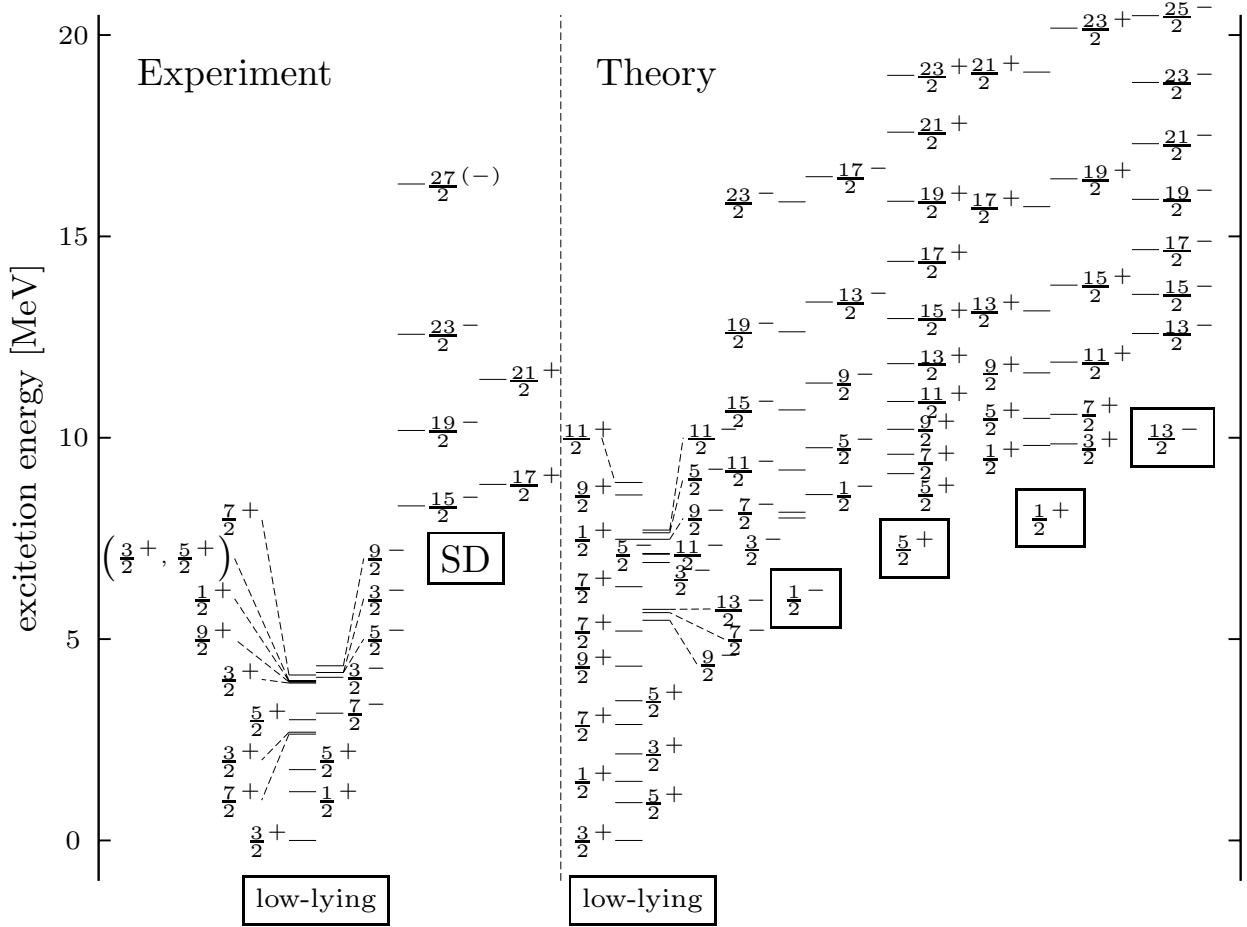}
\caption{
 Level scheme and of $^{35}$Cl.
 Left and right parts show experimental and theoretical levels.
 Experimental data are taken from Refs.~\onlinecite{PhysRevC.88.034303} and \onlinecite{CHEN20112715}.
 Labels of bands are shown in squares.
 }
\label{level_35Cl}
\end{figure*}

Figure~\ref{level_35Cl} shows level scheme of $^{35}$Cl obtained by the GCM.
Various rotational bands labeled $K^\pi = \frac{1}{2}^\pm$, $\frac{5}{2}^+$, and $\frac{13}{2}^-$ bands, are obtained as well as low-lying states.
The experimental states, the negative-parity SD band and candidates of parity-doublet partners, labeled SD, are also plotted as well as low-lying states.
Dominant components of the $K^\pi = \frac{1}{2}^-$ and $\frac{13}{2}^-$ bands have $3\hbar\omega$ excited configurations, and those of $K^\pi = \frac{1}{2}^+$ and $\frac{5}{2}^+$ bands have $4\hbar\omega$ excited configurations. 
Other low-lying state have $0\hbar\omega$ and $1\hbar\omega$ excited configurations for positive- and negative-parity states, respectively.
The obtained $J^\pi = \frac{19}{2}^-$ and $\frac{23}{2}^-$ states in the $K^\pi = \frac{1}{2}^-$ band are negative-parity yrast states.
In the $K^\pi = \frac{1}{2}^-$ band, a $J^\pi = \frac{3}{2}^-$ state has the lowest energy due to a large decoupling parameter.

\begin{figure}[tbp]
 \includegraphics[width=0.5\textwidth]{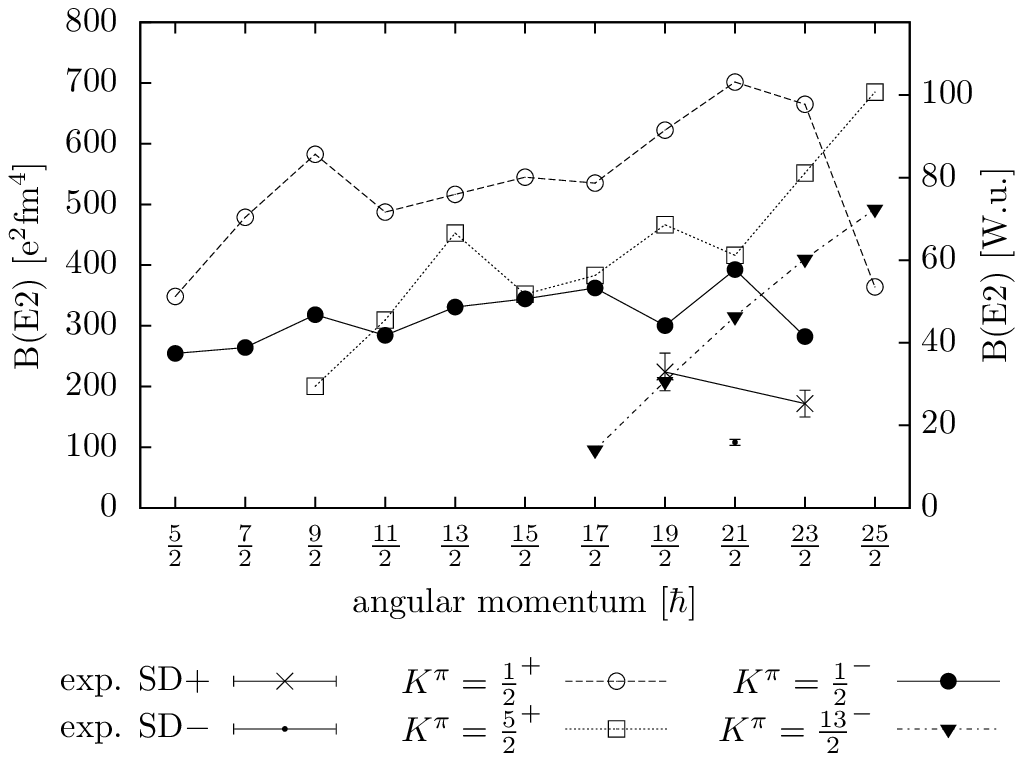}
 \caption{
 In-band $B(\textrm{E2}; J \rightarrow J - 2)$ values are plotted as functions of spin $J$ of an initial states for theoretical $K^\pi = \frac{1}{2}^\pm$, $\frac{5}{2}^+$, and $\frac{13}{2}^-$ bands and the experimental negative-parity SD band (SD$-$) and its candidate positive-parity partner band (SD+).
 Left and right vertical axes are in $\textrm{e}^2 \textrm{fm}^4$ and Weiskkopf unit, respectively.
 } 
 \label{BE2}
\end{figure} 


Figure~\ref{BE2} shows in-band B(E2; $J \rightarrow J - 2$) values for obtained rotational bands, where $J$ is a spin of an initial state.
Experimental values for the negative-parity SD band and its candidates of parity partner states are also shown.
In-band $B(\mathrm{E2})$ values of the $K = \frac{1}{2}^-$ band are consistent with those of the experimental negative-parity SD band.
In-band $B(\mathrm{E2})$ values of the $K = \frac{1}{2}^-$ band are 40--60~W.u., and those of the experimental negative-parity SD band are $32.9\pm 4.6$~W.u. and $25.3\pm 3.2$~W.u. for $\frac{19}{2}^-\rightarrow\frac{15}{2}^-$ and $\frac{23}{2}^-\rightarrow\frac{19}{2}^-$ transitions, respectively.

\begin{figure}[tbp]
 \includegraphics[width=0.5\textwidth]{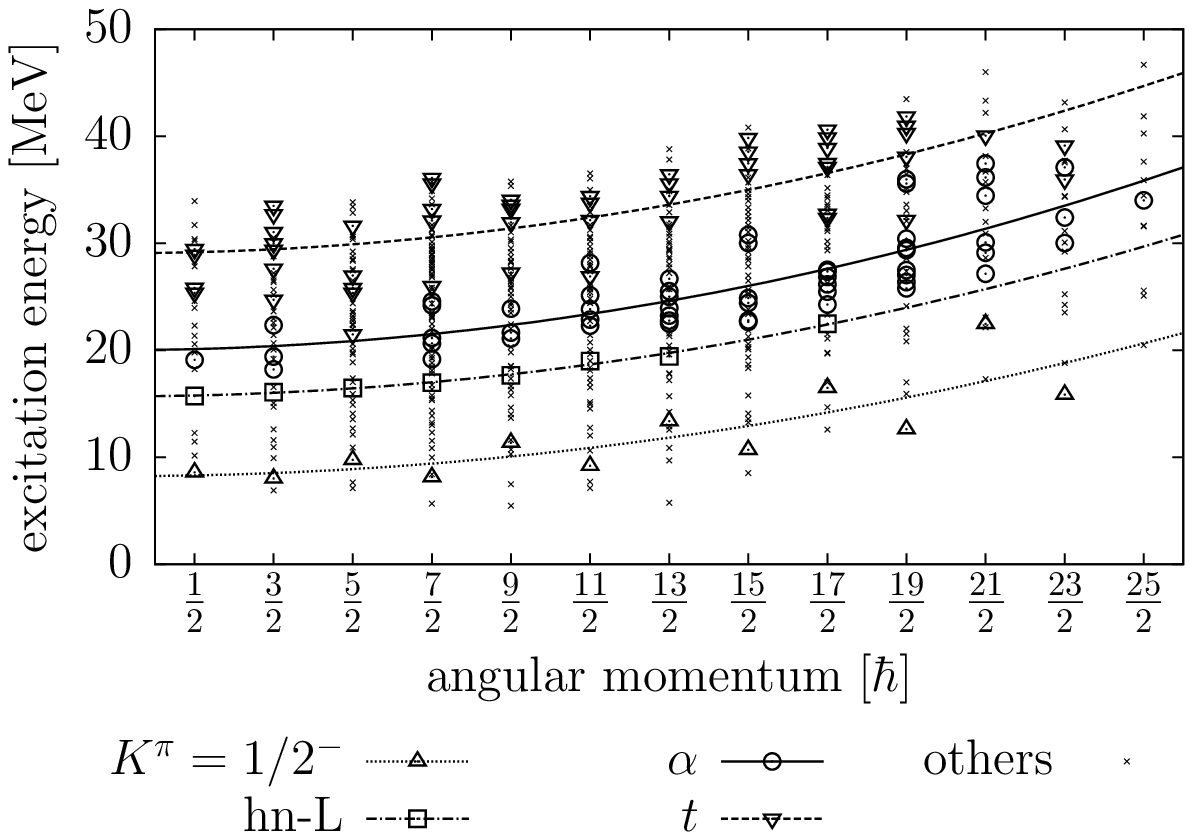}
 \caption{
 Excitation energies of obtained negative-parity states are plotted.
 Horizontal axis shows spin of each state.
 Triangle, squares, circles, and inverted triangles show members of the $K^\pi = \frac{1}{2}^-$, hn-L, ``$\alpha$'', and ``$t$'' bands, respectively (see text).
 Crosses shows shows other states.
 Dotted, chain, solid, and dashed lines show trends of the excitation energies of the $K^\pi = \frac{1}{2}^-$, hn-L, ``$\alpha$'', and ``$t$'' bands, respectively.
 }
 \label{cluster}
\end{figure}

Figure~\ref{cluster} shows excitation energies of obtained negative-parity states.
States that contains large amounts of L-type $\alpha$-$^{31}$P and $t$-$^{32}$S cluster structure components are stressed.
Many states contain large amounts of L-type cluster structure components in high-lying states as well as low-lying $K = \frac{1}{2}^-$ bands.
Higher-nodal states of L-type cluster structures (hn-L) are also obtained.
In high-lying states, almost pure $\alpha$-$^{31}$P and $t$-$^{32}$S cluster states are obtained, which are labeled ``$\alpha$'' and ``$t$'', respectively, following.

In order to analyze trend of excitation energies of the $K^\pi = \frac{1}{2}^-$, hn-L, ``$\alpha$'', and ``$t$'' bands, excitation energies of those states are fitted to following function as shown in Fig.~\ref{cluster};
\begin{equation}
 E(J) = \frac{J(J+1)}{2\mathcal{J}} + E_0,
\end{equation}
where $E$ and $J$ denote excitation energy and spin, respectively, and $\mathcal{J}$ and $E_0$ are fitting parameters to denote moment of inertia and energy for structure changes, respectively.
Excitation energies of the ``$t$'' band are approximately 10~MeV higher than those of the ``$\alpha$'' band.

\begin{figure*}[tbp]
 \begin{tabular}{cc}
  \includegraphics[width=0.45\textwidth]{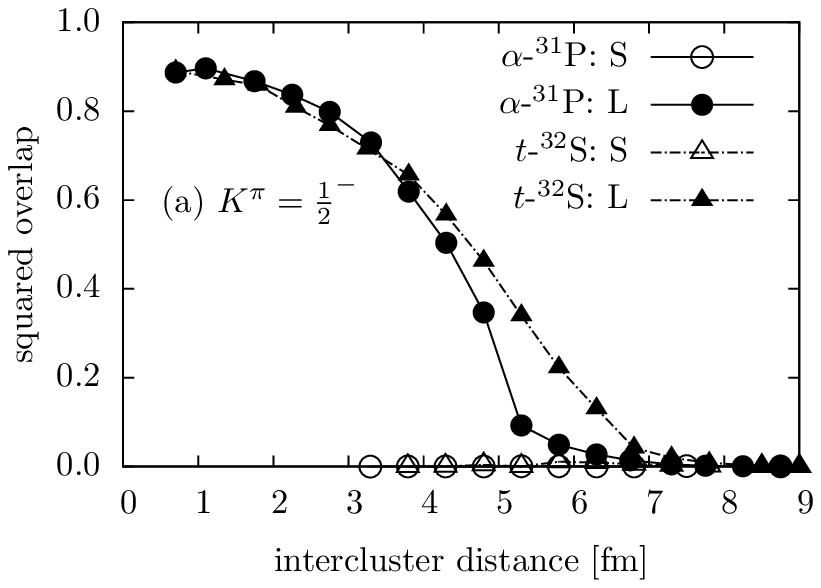}&
  \includegraphics[width=0.45\textwidth]{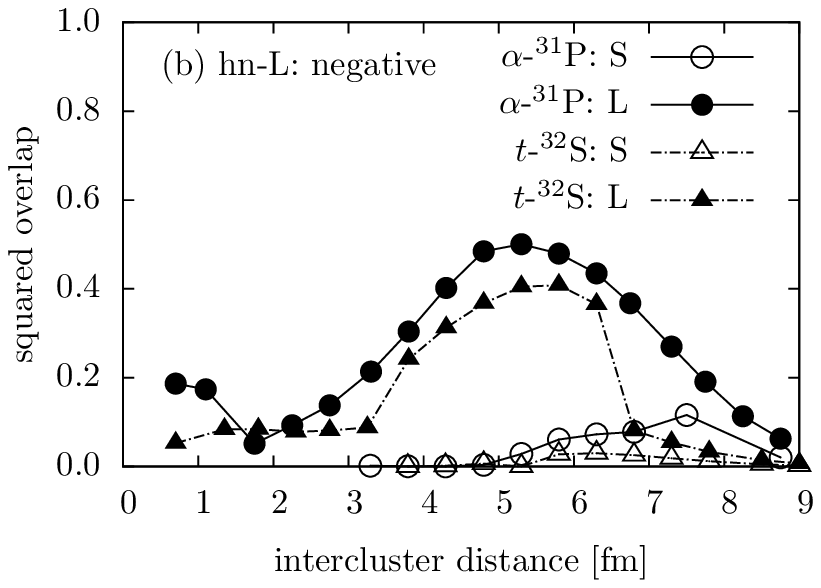}\\
  \includegraphics[width=0.45\textwidth]{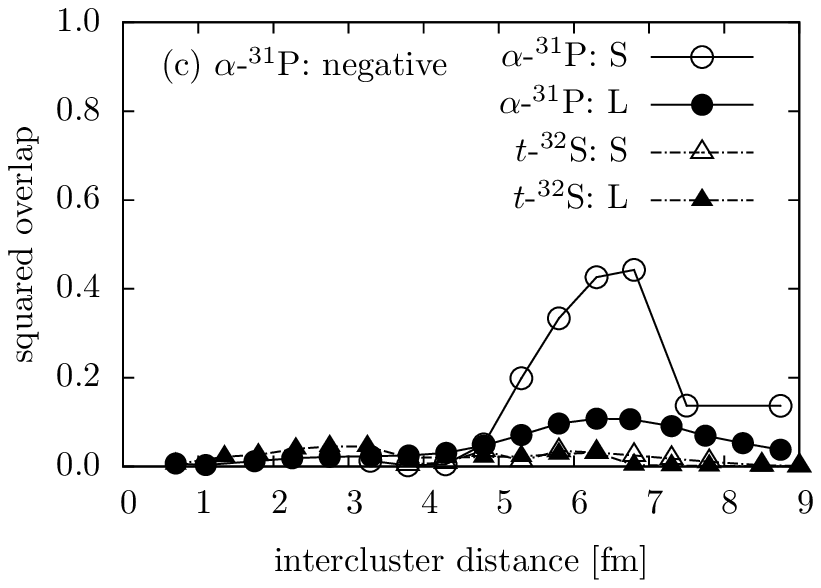}&
  \includegraphics[width=0.45\textwidth]{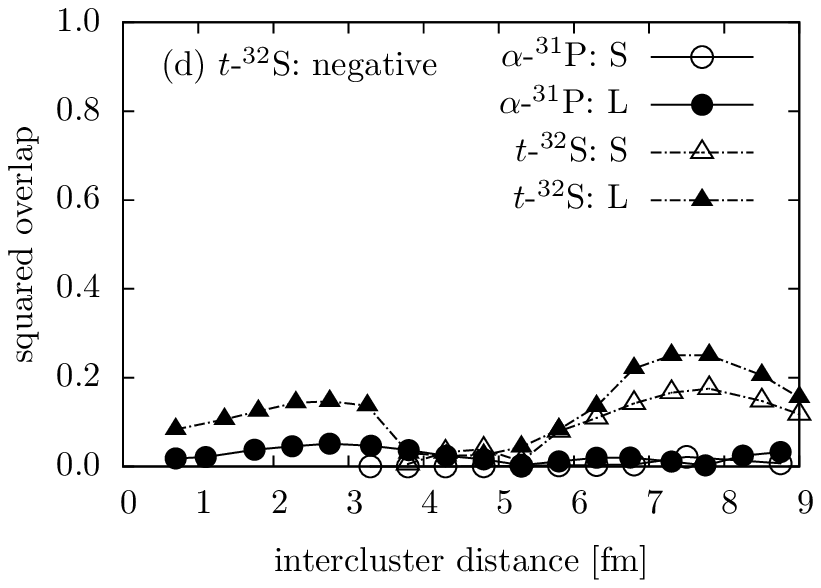}\\
 \end{tabular}
\caption{
 Cluster structure components of members of the (a) $K = \frac{1}{2}^-$, (b) hn-L, (c) ``$\alpha$'', and (d) ``$t$'' bands.
 Solid lines with circles and dashed lines with triangles are for $\alpha$-$^{31}$P and $t$-$^{32}$S cluster structure components, respectively.
 Open and closed symbols denote S- and L-types, respectively.
 }
\label{cluster_structure}
\end{figure*}

Figure~\ref{cluster_structure} shows \AP and \tS cluster structure components of $J^- = \frac{3}{2}^-$ states in the $K = \frac{1}{2}^-$, hn-L, ``$\alpha$'', and ``$t$'' bands. 
For $K^\pi = \frac{1}{2}^-$ band, the L-type $\alpha$-$^{31}$P and $t$-$^{32}$S cluster structure components are similar for all intercluster distance region [Fig.~\ref{cluster_structure}(a)].
In short distance region, overlaps are more than 0.8, which means that those wave functions are also dominant components of the band as well as $3\hbar\omega$ excited deformed structure.
With increasing intercluster distance, the overlaps are decreasing, but the overlaps have long tail distribution.
Overlaps are still nonneglegible in surface region, $\dboth \gtrsim 4~\mathrm{fm}$.
S-type components are not contained.
In hn-L states, L-type \AP and \tS cluster structure components with $\dboth\sim 5$--6~fm are dominated as shown in Fig.~\ref{cluster_structure}(b), and short distance components are suppressed.
Figures~\ref{cluster_structure}(c) and (d) show cluster structure components in the ``$\alpha$'' and ``$t$'' bands, respectively.
The ``$\alpha$'' and ``$t$'' states contain large amount of \AP and \tS cluster structure components, respectively, with large intercluster distance region.
The ``$\alpha$'' and ``$t$'' bands contain both L- and S-type components, and \AP and \tS cluster structure components are decoupled.
Deformed structure components are rarely contained in the hn-L, ``$\alpha$'' and ``$t$'' states.

\section{Discussions}
\label{discussions}

The experimental $J^\pi = \frac{15}{2}^-$ (8.31~MeV), $\frac{19}{2}^-$ (10.18~MeV), and $\frac{23}{2}^-$ (12.57~MeV) states, which are assigned to a negative-parity SD band\cite{PhysRevC.88.034303}, correspond to members of the theoretical $K^\pi = \frac{1}{2}^-$ band.
The theoretical $J^\pi = \frac{19}{2}^-$ and $\frac{23}{2}^-$ states in the $K^\pi = \frac{1}{2}^-$ band are negative-parity yrast states, and those in the experimental negative-parity SD band are also yrast states.
In-band $B(\textrm{E2})$ values of the theoretical $K^\pi = \frac{1}{2}^-$ band and the experimental negative-parity SD band are consistent (Fig.~\ref{BE2}) although theoretical values are slightly overestimated.
Dominant components of the $K^\pi = \frac{1}{2}^-$ band have $3\hbar\omega$ excited configurations, which is consistent with a shell-model calculation\cite{PhysRevC.88.034303}.

In coupling of cluster structure components for relatively low energy states, particle-hole configuration of cluster wave functions with short intercluster distance are important rather than threshold energies.
Members of the $K^\pi = \frac{1}{2}^-$ band contain similar amounts of L-type \AP and \tS cluster structure components with short intercluster distance region as shown in Fig.~\ref{cluster_structure}(a).
Cluster and deformed structure components dominate in the $K^\pi = \frac{1}{2}^-$ that has same particle-hole configurations, which are $3\hbar\omega$ excited configurations.
Therefore, overlap between $3\hbar\omega$ excited deformed structure and L-type cluster structures with short intercluster distance is large, and coupling and mixing of those structures are strong in the $K^\pi = \frac{1}{2}^-$ band.
Configurations of cluster structure change gradually with increasing intercluster distance, and the $K^\pi = \frac{1}{2}^-$ band contain L-type cluster structure components up to tail region.
The similar strength of \AP and \tS cluster coupling in the $K^\pi = \frac{1}{2}^-$ bands is inconsistent with the threshold energy rule because threshold energies of $\alpha + {}^{31}\textrm{P}$ (6.99~MeV) and $t + {}^{32}\textrm{S}$ (17.94~MeV) are much different.

The hn-L states are generated by excitation of intercluster motion of cluster structure components in the $K^\pi = \frac{1}{2}^-$ band.
Dominant components of the hn-L states are L-type \AP and \tS cluster structure components with $\dAP$ and $\dtS\sim 5$~fm, respectively [Fig.~\ref{cluster_structure}(b)]. 
Compared to L-type \AP and \tS cluster structure components contained in the $K^\pi = \frac{1}{2}^-$ band, 
cluster structure components with short intercluster distance are suppressed in the hn-L states.
The suppression is caused by orthogonalization to the $K^\pi = \frac{1}{2}^-$ band. 
Deformed structure components are rarely contained in the hn-L states.
They show that hn-L states are generated by excitation of cluster structure components in the $K^\pi = \frac{1}{2}^-$ band.

For cluster structures in high-lying states, threshold energies reflect to excitation energies and correlation between clusters are weak.
In negative-parity states, almost pure \AP and \tS cluster states are obtained, which are labeled ``$\alpha$'' and ``$t$'' in Fig.~\ref{cluster}, respectively.
Excitation energies of \tS cluster states are approximately 10~MeV higher than those of \AP cluster states.
The energy gap is similar to difference of threshold energies of $\alpha + {}^{31}\textrm{P}$ and $t + {}^{32}\textrm{S}$.
Intercluster distance of dominant components of the ``$\alpha$'' and ``$t$'' states are elongated to Coulomb barrier region [Figs.~\ref{cluster_structure}(c) and (d)] by orthogonalization to the hn-L states, which shows excitation of intercluster motion of cluster structure components in the hn-L states.
In long intercluster distance region, interaction between clusters become weaker.
Therefore, threshold energies strongly affect excitation energies, and coupling between clusters become weaker. 
Coupling between clusters in ``$\alpha$'' and ``$t$'' states are weaker than that of $K^\pi = \frac{1}{2}^-$ bands and  hn-L states.
Mixing of L- and S-type cluster structure components in ``$\alpha$'' and ``$t$'' states describes rotation of larger clusters due to weak correlation between clusters. 

\section{Conclusions}
\label{conclusions}

\AP and \tS clustering in $^{35}$Cl has been investigated by using the AMD and GCM.
The experimental negative-parity SD band is reproduced, and the SD band has $3\hbar\omega$ excited configurations.
The negative-parity SD band contains L-type \AP and \tS cluster structure components with short intercluster distance although threshold energies of $\alpha + {}^{31}\textrm{P}$ and $t + {}^{32}\textrm{S}$ are much different.
The mixing of L-type \AP and \tS cluster structure components in the negative-parity SD band are caused by coincidence of particle-hole configurations of cluster structure components with short intercluster distance and dominant $3\hbar\omega$ excited deformed structures.
In high-lying states, almost pure \tS and \AP states are obtained in negative-parity states.
Threshold energies reflect to excitation energies of those almost pure cluster states.
By excitation of intercluster motion, cluster structure components with long intercluster distance are dominated in high-lying cluster states, in which threshold energies reflect to total energies.
In coupling of cluster structure components, particle-hole configurations in short intercluster distance region and threshold energies are important for low- and high-lying states, respectively.

\begin{acknowledgments}
The author thanks to Dr.~Kimura in Hokkaido University for fruitful discussions.
This work was supported by JSPS KAKENHI Grant Number 25800124, Multidisciplinary Cooperative Research Program in CCS, University of Tsukuba, and The Hattori Hokokai Foundation
Grant-in-Aid for Technological and Engineering Research.
\end{acknowledgments}

\bibliography{35Cl_v1}

\begin{thebibliography}{42}%
\makeatletter
\providecommand \@ifxundefined [1]{%
 \@ifx{#1\undefined}
}%
\providecommand \@ifnum [1]{%
 \ifnum #1\expandafter \@firstoftwo
 \else \expandafter \@secondoftwo
 \fi
}%
\providecommand \@ifx [1]{%
 \ifx #1\expandafter \@firstoftwo
 \else \expandafter \@secondoftwo
 \fi
}%
\providecommand \natexlab [1]{#1}%
\providecommand \enquote  [1]{``#1''}%
\providecommand \bibnamefont  [1]{#1}%
\providecommand \bibfnamefont [1]{#1}%
\providecommand \citenamefont [1]{#1}%
\providecommand \href@noop [0]{\@secondoftwo}%
\providecommand \href [0]{\begingroup \@sanitize@url \@href}%
\providecommand \@href[1]{\@@startlink{#1}\@@href}%
\providecommand \@@href[1]{\endgroup#1\@@endlink}%
\providecommand \@sanitize@url [0]{\catcode `\\12\catcode `\$12\catcode
  `\&12\catcode `\#12\catcode `\^12\catcode `\_12\catcode `\%12\relax}%
\providecommand \@@startlink[1]{}%
\providecommand \@@endlink[0]{}%
\providecommand \url  [0]{\begingroup\@sanitize@url \@url }%
\providecommand \@url [1]{\endgroup\@href {#1}{\urlprefix }}%
\providecommand \urlprefix  [0]{URL }%
\providecommand \Eprint [0]{\href }%
\providecommand \doibase [0]{http://dx.doi.org/}%
\providecommand \selectlanguage [0]{\@gobble}%
\providecommand \bibinfo  [0]{\@secondoftwo}%
\providecommand \bibfield  [0]{\@secondoftwo}%
\providecommand \translation [1]{[#1]}%
\providecommand \BibitemOpen [0]{}%
\providecommand \bibitemStop [0]{}%
\providecommand \bibitemNoStop [0]{.\EOS\space}%
\providecommand \EOS [0]{\spacefactor3000\relax}%
\providecommand \BibitemShut  [1]{\csname bibitem#1\endcsname}%
\let\auto@bib@innerbib\@empty
\bibitem [{\citenamefont {Horiuchi}\ and\ \citenamefont
  {Ikeda}(1968)}]{10.1143/PTP.40.277}%
  \BibitemOpen
  \bibfield  {author} {\bibinfo {author} {\bibfnamefont {H.}~\bibnamefont
  {Horiuchi}}\ and\ \bibinfo {author} {\bibfnamefont {K.}~\bibnamefont
  {Ikeda}},\ }\href {\doibase 10.1143/PTP.40.277} {\bibfield  {journal}
  {\bibinfo  {journal} {Progress of Theoretical Physics}\ }\textbf {\bibinfo
  {volume} {40}},\ \bibinfo {pages} {277} (\bibinfo {year} {1968})}\BibitemShut
  {NoStop}%
\bibitem [{\citenamefont {Becchetti}\ \emph {et~al.}(1978)\citenamefont
  {Becchetti}, \citenamefont {Jänecke},\ and\ \citenamefont
  {Thorn}}]{BECCHETTI1978313}%
  \BibitemOpen
  \bibfield  {author} {\bibinfo {author} {\bibfnamefont {F.}~\bibnamefont
  {Becchetti}}, \bibinfo {author} {\bibfnamefont {J.}~\bibnamefont {Jänecke}},
  \ and\ \bibinfo {author} {\bibfnamefont {C.}~\bibnamefont {Thorn}},\ }\href
  {\doibase https://doi.org/10.1016/0375-9474(78)90180-X} {\bibfield  {journal}
  {\bibinfo  {journal} {Nuclear Physics A}\ }\textbf {\bibinfo {volume}
  {305}},\ \bibinfo {pages} {313 } (\bibinfo {year} {1978})}\BibitemShut
  {NoStop}%
\bibitem [{\citenamefont {Anantaraman}\ \emph {et~al.}(1979)\citenamefont
  {Anantaraman}, \citenamefont {Gove}, \citenamefont {Lindgren}, \citenamefont
  {Tōke}, \citenamefont {Trentelman}, \citenamefont {Draayer}, \citenamefont
  {Jundt},\ and\ \citenamefont {Guillaume}}]{ANANTARAMAN1979445}%
  \BibitemOpen
  \bibfield  {author} {\bibinfo {author} {\bibfnamefont {N.}~\bibnamefont
  {Anantaraman}}, \bibinfo {author} {\bibfnamefont {H.}~\bibnamefont {Gove}},
  \bibinfo {author} {\bibfnamefont {R.}~\bibnamefont {Lindgren}}, \bibinfo
  {author} {\bibfnamefont {J.}~\bibnamefont {Tōke}}, \bibinfo {author}
  {\bibfnamefont {J.}~\bibnamefont {Trentelman}}, \bibinfo {author}
  {\bibfnamefont {J.}~\bibnamefont {Draayer}}, \bibinfo {author} {\bibfnamefont
  {F.}~\bibnamefont {Jundt}}, \ and\ \bibinfo {author} {\bibfnamefont
  {G.}~\bibnamefont {Guillaume}},\ }\href {\doibase
  https://doi.org/10.1016/0375-9474(79)90512-8} {\bibfield  {journal} {\bibinfo
   {journal} {Nuclear Physics A}\ }\textbf {\bibinfo {volume} {313}},\ \bibinfo
  {pages} {445 } (\bibinfo {year} {1979})}\BibitemShut {NoStop}%
\bibitem [{\citenamefont {Fukui}\ \emph {et~al.}(2016)\citenamefont {Fukui},
  \citenamefont {Taniguchi}, \citenamefont {Suhara}, \citenamefont
  {Kanada-En'yo},\ and\ \citenamefont {Ogata}}]{PhysRevC.93.034606}%
  \BibitemOpen
  \bibfield  {author} {\bibinfo {author} {\bibfnamefont {T.}~\bibnamefont
  {Fukui}}, \bibinfo {author} {\bibfnamefont {Y.}~\bibnamefont {Taniguchi}},
  \bibinfo {author} {\bibfnamefont {T.}~\bibnamefont {Suhara}}, \bibinfo
  {author} {\bibfnamefont {Y.}~\bibnamefont {Kanada-En'yo}}, \ and\ \bibinfo
  {author} {\bibfnamefont {K.}~\bibnamefont {Ogata}},\ }\href {\doibase
  10.1103/PhysRevC.93.034606} {\bibfield  {journal} {\bibinfo  {journal} {Phys.
  Rev. C}\ }\textbf {\bibinfo {volume} {93}},\ \bibinfo {pages} {034606}
  (\bibinfo {year} {2016})}\BibitemShut {NoStop}%
\bibitem [{\citenamefont {Fukui}\ \emph {et~al.}(2019)\citenamefont {Fukui},
  \citenamefont {Kanada-En'yo}, \citenamefont {Ogata}, \citenamefont {Suhara},\
  and\ \citenamefont {Taniguchi}}]{FUKUI201938}%
  \BibitemOpen
  \bibfield  {author} {\bibinfo {author} {\bibfnamefont {T.}~\bibnamefont
  {Fukui}}, \bibinfo {author} {\bibfnamefont {Y.}~\bibnamefont {Kanada-En'yo}},
  \bibinfo {author} {\bibfnamefont {K.}~\bibnamefont {Ogata}}, \bibinfo
  {author} {\bibfnamefont {T.}~\bibnamefont {Suhara}}, \ and\ \bibinfo {author}
  {\bibfnamefont {Y.}~\bibnamefont {Taniguchi}},\ }\href {\doibase
  https://doi.org/10.1016/j.nuclphysa.2018.12.024} {\bibfield  {journal}
  {\bibinfo  {journal} {Nuclear Physics A}\ }\textbf {\bibinfo {volume}
  {983}},\ \bibinfo {pages} {38 } (\bibinfo {year} {2019})}\BibitemShut
  {NoStop}%
\bibitem [{\citenamefont {Roos}\ \emph {et~al.}(1977)\citenamefont {Roos},
  \citenamefont {Chant}, \citenamefont {Cowley}, \citenamefont {Goldberg},
  \citenamefont {Holmgren},\ and\ \citenamefont {Woody}}]{PhysRevC.15.69}%
  \BibitemOpen
  \bibfield  {author} {\bibinfo {author} {\bibfnamefont {P.~G.}\ \bibnamefont
  {Roos}}, \bibinfo {author} {\bibfnamefont {N.~S.}\ \bibnamefont {Chant}},
  \bibinfo {author} {\bibfnamefont {A.~A.}\ \bibnamefont {Cowley}}, \bibinfo
  {author} {\bibfnamefont {D.~A.}\ \bibnamefont {Goldberg}}, \bibinfo {author}
  {\bibfnamefont {H.~D.}\ \bibnamefont {Holmgren}}, \ and\ \bibinfo {author}
  {\bibfnamefont {R.}~\bibnamefont {Woody}},\ }\href {\doibase
  10.1103/PhysRevC.15.69} {\bibfield  {journal} {\bibinfo  {journal} {Phys.
  Rev. C}\ }\textbf {\bibinfo {volume} {15}},\ \bibinfo {pages} {69} (\bibinfo
  {year} {1977})}\BibitemShut {NoStop}%
\bibitem [{\citenamefont {Nadasen}\ \emph {et~al.}(1980)\citenamefont
  {Nadasen}, \citenamefont {Chant}, \citenamefont {Roos}, \citenamefont
  {Carey}, \citenamefont {Cowen}, \citenamefont {Samanta},\ and\ \citenamefont
  {Wesick}}]{PhysRevC.22.1394}%
  \BibitemOpen
  \bibfield  {author} {\bibinfo {author} {\bibfnamefont {A.}~\bibnamefont
  {Nadasen}}, \bibinfo {author} {\bibfnamefont {N.~S.}\ \bibnamefont {Chant}},
  \bibinfo {author} {\bibfnamefont {P.~G.}\ \bibnamefont {Roos}}, \bibinfo
  {author} {\bibfnamefont {T.~A.}\ \bibnamefont {Carey}}, \bibinfo {author}
  {\bibfnamefont {R.}~\bibnamefont {Cowen}}, \bibinfo {author} {\bibfnamefont
  {C.}~\bibnamefont {Samanta}}, \ and\ \bibinfo {author} {\bibfnamefont
  {J.}~\bibnamefont {Wesick}},\ }\href {\doibase 10.1103/PhysRevC.22.1394}
  {\bibfield  {journal} {\bibinfo  {journal} {Phys. Rev. C}\ }\textbf {\bibinfo
  {volume} {22}},\ \bibinfo {pages} {1394} (\bibinfo {year}
  {1980})}\BibitemShut {NoStop}%
\bibitem [{\citenamefont {Wang}\ \emph {et~al.}(1985)\citenamefont {Wang},
  \citenamefont {Roos}, \citenamefont {Chant}, \citenamefont {Ciangaru},
  \citenamefont {Khazaie}, \citenamefont {J.~Mack}, \citenamefont {Nadasen},
  \citenamefont {Mills}, \citenamefont {Warner}, \citenamefont {Norbeck},
  \citenamefont {Becchetti}, \citenamefont {Janecke},\ and\ \citenamefont
  {Lister}}]{PhysRevC.31.1662}%
  \BibitemOpen
  \bibfield  {author} {\bibinfo {author} {\bibfnamefont {C.~W.}\ \bibnamefont
  {Wang}}, \bibinfo {author} {\bibfnamefont {P.~G.}\ \bibnamefont {Roos}},
  \bibinfo {author} {\bibfnamefont {N.~S.}\ \bibnamefont {Chant}}, \bibinfo
  {author} {\bibfnamefont {G.}~\bibnamefont {Ciangaru}}, \bibinfo {author}
  {\bibfnamefont {F.}~\bibnamefont {Khazaie}}, \bibinfo {author} {\bibfnamefont
  {D.}~\bibnamefont {J.~Mack}}, \bibinfo {author} {\bibfnamefont
  {A.}~\bibnamefont {Nadasen}}, \bibinfo {author} {\bibfnamefont {S.~J.}\
  \bibnamefont {Mills}}, \bibinfo {author} {\bibfnamefont {R.~E.}\ \bibnamefont
  {Warner}}, \bibinfo {author} {\bibfnamefont {E.}~\bibnamefont {Norbeck}},
  \bibinfo {author} {\bibfnamefont {F.~D.}\ \bibnamefont {Becchetti}}, \bibinfo
  {author} {\bibfnamefont {J.~W.}\ \bibnamefont {Janecke}}, \ and\ \bibinfo
  {author} {\bibfnamefont {P.~M.}\ \bibnamefont {Lister}},\ }\href {\doibase
  10.1103/PhysRevC.31.1662} {\bibfield  {journal} {\bibinfo  {journal} {Phys.
  Rev. C}\ }\textbf {\bibinfo {volume} {31}},\ \bibinfo {pages} {1662}
  (\bibinfo {year} {1985})}\BibitemShut {NoStop}%
\bibitem [{\citenamefont {Nadasen}\ \emph {et~al.}(1989)\citenamefont
  {Nadasen}, \citenamefont {Roos}, \citenamefont {Chant}, \citenamefont
  {Chang}, \citenamefont {Ciangaru}, \citenamefont {Breuer}, \citenamefont
  {Wesick},\ and\ \citenamefont {Norbeck}}]{PhysRevC.40.1130}%
  \BibitemOpen
  \bibfield  {author} {\bibinfo {author} {\bibfnamefont {A.}~\bibnamefont
  {Nadasen}}, \bibinfo {author} {\bibfnamefont {P.~G.}\ \bibnamefont {Roos}},
  \bibinfo {author} {\bibfnamefont {N.~S.}\ \bibnamefont {Chant}}, \bibinfo
  {author} {\bibfnamefont {C.~C.}\ \bibnamefont {Chang}}, \bibinfo {author}
  {\bibfnamefont {G.}~\bibnamefont {Ciangaru}}, \bibinfo {author}
  {\bibfnamefont {H.~F.}\ \bibnamefont {Breuer}}, \bibinfo {author}
  {\bibfnamefont {J.}~\bibnamefont {Wesick}}, \ and\ \bibinfo {author}
  {\bibfnamefont {E.}~\bibnamefont {Norbeck}},\ }\href {\doibase
  10.1103/PhysRevC.40.1130} {\bibfield  {journal} {\bibinfo  {journal} {Phys.
  Rev. C}\ }\textbf {\bibinfo {volume} {40}},\ \bibinfo {pages} {1130}
  (\bibinfo {year} {1989})}\BibitemShut {NoStop}%
\bibitem [{\citenamefont {Yoshida}\ \emph {et~al.}(2018)\citenamefont
  {Yoshida}, \citenamefont {Ogata},\ and\ \citenamefont
  {Kanada-En'yo}}]{PhysRevC.98.024614}%
  \BibitemOpen
  \bibfield  {author} {\bibinfo {author} {\bibfnamefont {K.}~\bibnamefont
  {Yoshida}}, \bibinfo {author} {\bibfnamefont {K.}~\bibnamefont {Ogata}}, \
  and\ \bibinfo {author} {\bibfnamefont {Y.}~\bibnamefont {Kanada-En'yo}},\
  }\href {\doibase 10.1103/PhysRevC.98.024614} {\bibfield  {journal} {\bibinfo
  {journal} {Phys. Rev. C}\ }\textbf {\bibinfo {volume} {98}},\ \bibinfo
  {pages} {024614} (\bibinfo {year} {2018})}\BibitemShut {NoStop}%
\bibitem [{\citenamefont {Ideguchi}\ \emph {et~al.}(2001)\citenamefont
  {Ideguchi}, \citenamefont {Sarantites}, \citenamefont {Reviol}, \citenamefont
  {Afanasjev}, \citenamefont {Devlin}, \citenamefont {Baktash}, \citenamefont
  {Janssens}, \citenamefont {Rudolph}, \citenamefont {Axelsson}, \citenamefont
  {Carpenter}, \citenamefont {Galindo-Uribarri}, \citenamefont {LaFosse},
  \citenamefont {Lauritsen}, \citenamefont {Lerma}, \citenamefont {Lister},
  \citenamefont {Reiter}, \citenamefont {Seweryniak}, \citenamefont
  {Weiszflog},\ and\ \citenamefont {Wilson}}]{PhysRevLett.87.222501}%
  \BibitemOpen
  \bibfield  {author} {\bibinfo {author} {\bibfnamefont {E.}~\bibnamefont
  {Ideguchi}}, \bibinfo {author} {\bibfnamefont {D.~G.}\ \bibnamefont
  {Sarantites}}, \bibinfo {author} {\bibfnamefont {W.}~\bibnamefont {Reviol}},
  \bibinfo {author} {\bibfnamefont {A.~V.}\ \bibnamefont {Afanasjev}}, \bibinfo
  {author} {\bibfnamefont {M.}~\bibnamefont {Devlin}}, \bibinfo {author}
  {\bibfnamefont {C.}~\bibnamefont {Baktash}}, \bibinfo {author} {\bibfnamefont
  {R.~V.~F.}\ \bibnamefont {Janssens}}, \bibinfo {author} {\bibfnamefont
  {D.}~\bibnamefont {Rudolph}}, \bibinfo {author} {\bibfnamefont
  {A.}~\bibnamefont {Axelsson}}, \bibinfo {author} {\bibfnamefont {M.~P.}\
  \bibnamefont {Carpenter}}, \bibinfo {author} {\bibfnamefont {A.}~\bibnamefont
  {Galindo-Uribarri}}, \bibinfo {author} {\bibfnamefont {D.~R.}\ \bibnamefont
  {LaFosse}}, \bibinfo {author} {\bibfnamefont {T.}~\bibnamefont {Lauritsen}},
  \bibinfo {author} {\bibfnamefont {F.}~\bibnamefont {Lerma}}, \bibinfo
  {author} {\bibfnamefont {C.~J.}\ \bibnamefont {Lister}}, \bibinfo {author}
  {\bibfnamefont {P.}~\bibnamefont {Reiter}}, \bibinfo {author} {\bibfnamefont
  {D.}~\bibnamefont {Seweryniak}}, \bibinfo {author} {\bibfnamefont
  {M.}~\bibnamefont {Weiszflog}}, \ and\ \bibinfo {author} {\bibfnamefont
  {J.~N.}\ \bibnamefont {Wilson}},\ }\href {\doibase
  10.1103/PhysRevLett.87.222501} {\bibfield  {journal} {\bibinfo  {journal}
  {Phys. Rev. Lett.}\ }\textbf {\bibinfo {volume} {87}},\ \bibinfo {pages}
  {222501} (\bibinfo {year} {2001})}\BibitemShut {NoStop}%
\bibitem [{\citenamefont {Ohkubo}\ and\ \citenamefont
  {Umehara}(1988)}]{PTP.80.598}%
  \BibitemOpen
  \bibfield  {author} {\bibinfo {author} {\bibfnamefont {S.}~\bibnamefont
  {Ohkubo}}\ and\ \bibinfo {author} {\bibfnamefont {K.}~\bibnamefont
  {Umehara}},\ }\href {\doibase 10.1143/PTP.80.598} {\bibfield  {journal}
  {\bibinfo  {journal} {Progress of Theoretical Physics}\ }\textbf {\bibinfo
  {volume} {80}},\ \bibinfo {pages} {598} (\bibinfo {year} {1988})}\BibitemShut
  {NoStop}%
\bibitem [{\citenamefont {Reidemeister}\ \emph {et~al.}(1990)\citenamefont
  {Reidemeister}, \citenamefont {Ohkubo},\ and\ \citenamefont
  {Michel}}]{PhysRevC.41.63}%
  \BibitemOpen
  \bibfield  {author} {\bibinfo {author} {\bibfnamefont {G.}~\bibnamefont
  {Reidemeister}}, \bibinfo {author} {\bibfnamefont {S.}~\bibnamefont
  {Ohkubo}}, \ and\ \bibinfo {author} {\bibfnamefont {F.}~\bibnamefont
  {Michel}},\ }\href {\doibase 10.1103/PhysRevC.41.63} {\bibfield  {journal}
  {\bibinfo  {journal} {Phys. Rev. C}\ }\textbf {\bibinfo {volume} {41}},\
  \bibinfo {pages} {63} (\bibinfo {year} {1990})}\BibitemShut {NoStop}%
\bibitem [{\citenamefont {Yamaya}\ \emph {et~al.}(1994)\citenamefont {Yamaya},
  \citenamefont {Saitoh}, \citenamefont {Fujiwara}, \citenamefont {Itahashi},
  \citenamefont {Katori}, \citenamefont {Suehiro}, \citenamefont {Kato},
  \citenamefont {Hatori},\ and\ \citenamefont {Ohkubo}}]{Yamaya1994154}%
  \BibitemOpen
  \bibfield  {author} {\bibinfo {author} {\bibfnamefont {T.}~\bibnamefont
  {Yamaya}}, \bibinfo {author} {\bibfnamefont {M.}~\bibnamefont {Saitoh}},
  \bibinfo {author} {\bibfnamefont {M.}~\bibnamefont {Fujiwara}}, \bibinfo
  {author} {\bibfnamefont {T.}~\bibnamefont {Itahashi}}, \bibinfo {author}
  {\bibfnamefont {K.}~\bibnamefont {Katori}}, \bibinfo {author} {\bibfnamefont
  {T.}~\bibnamefont {Suehiro}}, \bibinfo {author} {\bibfnamefont
  {S.}~\bibnamefont {Kato}}, \bibinfo {author} {\bibfnamefont {S.}~\bibnamefont
  {Hatori}}, \ and\ \bibinfo {author} {\bibfnamefont {S.}~\bibnamefont
  {Ohkubo}},\ }\href {\doibase DOI: 10.1016/0375-9474(94)90019-1} {\bibfield
  {journal} {\bibinfo  {journal} {Nuclear Physics A}\ }\textbf {\bibinfo
  {volume} {573}},\ \bibinfo {pages} {154 } (\bibinfo {year}
  {1994})}\BibitemShut {NoStop}%
\bibitem [{\citenamefont {Sakuda}\ and\ \citenamefont
  {Ohkubo}(1994)}]{PhysRevC.49.149}%
  \BibitemOpen
  \bibfield  {author} {\bibinfo {author} {\bibfnamefont {T.}~\bibnamefont
  {Sakuda}}\ and\ \bibinfo {author} {\bibfnamefont {S.}~\bibnamefont
  {Ohkubo}},\ }\href {\doibase 10.1103/PhysRevC.49.149} {\bibfield  {journal}
  {\bibinfo  {journal} {Phys. Rev. C}\ }\textbf {\bibinfo {volume} {49}},\
  \bibinfo {pages} {149} (\bibinfo {year} {1994})}\BibitemShut {NoStop}%
\bibitem [{\citenamefont {Taniguchi}\ \emph {et~al.}(2007)\citenamefont
  {Taniguchi}, \citenamefont {Kimura}, \citenamefont {Kanada-En'yo},\ and\
  \citenamefont {Horiuchi}}]{taniguchi:044317}%
  \BibitemOpen
  \bibfield  {author} {\bibinfo {author} {\bibfnamefont {Y.}~\bibnamefont
  {Taniguchi}}, \bibinfo {author} {\bibfnamefont {M.}~\bibnamefont {Kimura}},
  \bibinfo {author} {\bibfnamefont {Y.}~\bibnamefont {Kanada-En'yo}}, \ and\
  \bibinfo {author} {\bibfnamefont {H.}~\bibnamefont {Horiuchi}},\ }\href
  {\doibase 10.1103/PhysRevC.76.044317} {\bibfield  {journal} {\bibinfo
  {journal} {Physical Review C (Nuclear Physics)}\ }\textbf {\bibinfo {volume}
  {76}},\ \bibinfo {eid} {044317} (\bibinfo {year} {2007})}\BibitemShut
  {NoStop}%
\bibitem [{\citenamefont {Ohkubo}\ and\ \citenamefont
  {Yamashita}(2002)}]{PhysRevC.66.021301}%
  \BibitemOpen
  \bibfield  {author} {\bibinfo {author} {\bibfnamefont {S.}~\bibnamefont
  {Ohkubo}}\ and\ \bibinfo {author} {\bibfnamefont {K.}~\bibnamefont
  {Yamashita}},\ }\href {\doibase 10.1103/PhysRevC.66.021301} {\bibfield
  {journal} {\bibinfo  {journal} {Phys. Rev. C}\ }\textbf {\bibinfo {volume}
  {66}},\ \bibinfo {pages} {021301} (\bibinfo {year} {2002})}\BibitemShut
  {NoStop}%
\bibitem [{\citenamefont {Kimura}\ and\ \citenamefont
  {Horiuchi}(2004)}]{PhysRevC.69.051304}%
  \BibitemOpen
  \bibfield  {author} {\bibinfo {author} {\bibfnamefont {M.}~\bibnamefont
  {Kimura}}\ and\ \bibinfo {author} {\bibfnamefont {H.}~\bibnamefont
  {Horiuchi}},\ }\href {\doibase 10.1103/PhysRevC.69.051304} {\bibfield
  {journal} {\bibinfo  {journal} {Phys. Rev. C}\ }\textbf {\bibinfo {volume}
  {69}},\ \bibinfo {pages} {051304} (\bibinfo {year} {2004})}\BibitemShut
  {NoStop}%
\bibitem [{\citenamefont {Sakuda}\ and\ \citenamefont
  {Ohkubo}(2004)}]{Sakuda200477}%
  \BibitemOpen
  \bibfield  {author} {\bibinfo {author} {\bibfnamefont {T.}~\bibnamefont
  {Sakuda}}\ and\ \bibinfo {author} {\bibfnamefont {S.}~\bibnamefont
  {Ohkubo}},\ }\href {\doibase
  http://dx.doi.org/10.1016/j.nuclphysa.2004.08.016} {\bibfield  {journal}
  {\bibinfo  {journal} {Nuclear Physics A}\ }\textbf {\bibinfo {volume}
  {744}},\ \bibinfo {pages} {77 } (\bibinfo {year} {2004})}\BibitemShut
  {NoStop}%
\bibitem [{\citenamefont {Sakuda}\ and\ \citenamefont
  {Ohkubo}(1995)}]{PhysRevC.51.586}%
  \BibitemOpen
  \bibfield  {author} {\bibinfo {author} {\bibfnamefont {T.}~\bibnamefont
  {Sakuda}}\ and\ \bibinfo {author} {\bibfnamefont {S.}~\bibnamefont
  {Ohkubo}},\ }\href {\doibase 10.1103/PhysRevC.51.586} {\bibfield  {journal}
  {\bibinfo  {journal} {Phys. Rev. C}\ }\textbf {\bibinfo {volume} {51}},\
  \bibinfo {pages} {586} (\bibinfo {year} {1995})}\BibitemShut {NoStop}%
\bibitem [{\citenamefont {Taniguchi}(2014)}]{Taniguchi01072014}%
  \BibitemOpen
  \bibfield  {author} {\bibinfo {author} {\bibfnamefont {Y.}~\bibnamefont
  {Taniguchi}},\ }\href {\doibase 10.1093/ptep/ptu086} {\bibfield  {journal}
  {\bibinfo  {journal} {Progress of Theoretical and Experimental Physics}\
  }\textbf {\bibinfo {volume} {2014}},\ \bibinfo {pages} {073D01} (\bibinfo
  {year} {2014})}\BibitemShut {NoStop}%
\bibitem [{\citenamefont {Friedrich}\ and\ \citenamefont
  {Langanke}(1975)}]{FRIEDRICH197547}%
  \BibitemOpen
  \bibfield  {author} {\bibinfo {author} {\bibfnamefont {H.}~\bibnamefont
  {Friedrich}}\ and\ \bibinfo {author} {\bibfnamefont {K.}~\bibnamefont
  {Langanke}},\ }\href {\doibase https://doi.org/10.1016/0375-9474(75)90600-4}
  {\bibfield  {journal} {\bibinfo  {journal} {Nuclear Physics A}\ }\textbf
  {\bibinfo {volume} {252}},\ \bibinfo {pages} {47 } (\bibinfo {year}
  {1975})}\BibitemShut {NoStop}%
\bibitem [{\citenamefont {Yamaya}\ \emph {et~al.}(1990)\citenamefont {Yamaya},
  \citenamefont {Oh-ami}, \citenamefont {Fujiwara}, \citenamefont {Itahashi},
  \citenamefont {Katori}, \citenamefont {Tosaki}, \citenamefont {Kato},
  \citenamefont {Hatori},\ and\ \citenamefont {Ohkubo}}]{PhysRevC.42.1935}%
  \BibitemOpen
  \bibfield  {author} {\bibinfo {author} {\bibfnamefont {T.}~\bibnamefont
  {Yamaya}}, \bibinfo {author} {\bibfnamefont {S.}~\bibnamefont {Oh-ami}},
  \bibinfo {author} {\bibfnamefont {M.}~\bibnamefont {Fujiwara}}, \bibinfo
  {author} {\bibfnamefont {T.}~\bibnamefont {Itahashi}}, \bibinfo {author}
  {\bibfnamefont {K.}~\bibnamefont {Katori}}, \bibinfo {author} {\bibfnamefont
  {M.}~\bibnamefont {Tosaki}}, \bibinfo {author} {\bibfnamefont
  {S.}~\bibnamefont {Kato}}, \bibinfo {author} {\bibfnamefont {S.}~\bibnamefont
  {Hatori}}, \ and\ \bibinfo {author} {\bibfnamefont {S.}~\bibnamefont
  {Ohkubo}},\ }\href {\doibase 10.1103/PhysRevC.42.1935} {\bibfield  {journal}
  {\bibinfo  {journal} {Phys. Rev. C}\ }\textbf {\bibinfo {volume} {42}},\
  \bibinfo {pages} {1935} (\bibinfo {year} {1990})}\BibitemShut {NoStop}%
\bibitem [{\citenamefont {Kimura}\ and\ \citenamefont
  {Horiuchi}(2006)}]{Kimura200658}%
  \BibitemOpen
  \bibfield  {author} {\bibinfo {author} {\bibfnamefont {M.}~\bibnamefont
  {Kimura}}\ and\ \bibinfo {author} {\bibfnamefont {H.}~\bibnamefont
  {Horiuchi}},\ }\href {\doibase DOI: 10.1016/j.nuclphysa.2005.12.006}
  {\bibfield  {journal} {\bibinfo  {journal} {Nuclear Physics A}\ }\textbf
  {\bibinfo {volume} {767}},\ \bibinfo {pages} {58 } (\bibinfo {year}
  {2006})}\BibitemShut {NoStop}%
\bibitem [{\citenamefont {Ikeda}\ \emph {et~al.}(1968)\citenamefont {Ikeda},
  \citenamefont {Takigawa},\ and\ \citenamefont
  {Horiuchi}}]{doi:10.1143/PTPS.E68.464}%
  \BibitemOpen
  \bibfield  {author} {\bibinfo {author} {\bibfnamefont {K.}~\bibnamefont
  {Ikeda}}, \bibinfo {author} {\bibfnamefont {N.}~\bibnamefont {Takigawa}}, \
  and\ \bibinfo {author} {\bibfnamefont {H.}~\bibnamefont {Horiuchi}},\ }\href
  {\doibase 10.1143/PTPS.E68.464} {\bibfield  {journal} {\bibinfo  {journal}
  {Progress of Theoretical Physics Supplement}\ }\textbf {\bibinfo {volume}
  {E68}},\ \bibinfo {pages} {464} (\bibinfo {year} {1968})}\BibitemShut
  {NoStop}%
\bibitem [{\citenamefont {Uegaki}\ \emph {et~al.}(1977)\citenamefont {Uegaki},
  \citenamefont {Okabe}, \citenamefont {Abe},\ and\ \citenamefont
  {Tanaka}}]{10.1143/PTP.57.1262}%
  \BibitemOpen
  \bibfield  {author} {\bibinfo {author} {\bibfnamefont {E.}~\bibnamefont
  {Uegaki}}, \bibinfo {author} {\bibfnamefont {S.}~\bibnamefont {Okabe}},
  \bibinfo {author} {\bibfnamefont {Y.}~\bibnamefont {Abe}}, \ and\ \bibinfo
  {author} {\bibfnamefont {H.}~\bibnamefont {Tanaka}},\ }\href {\doibase
  10.1143/PTP.57.1262} {\bibfield  {journal} {\bibinfo  {journal} {Progress of
  Theoretical Physics}\ }\textbf {\bibinfo {volume} {57}},\ \bibinfo {pages}
  {1262} (\bibinfo {year} {1977})}\BibitemShut {NoStop}%
\bibitem [{\citenamefont {Uegaki}\ \emph {et~al.}(1979)\citenamefont {Uegaki},
  \citenamefont {Abe}, \citenamefont {Okabe},\ and\ \citenamefont
  {Tanaka}}]{10.1143/PTP.62.1621}%
  \BibitemOpen
  \bibfield  {author} {\bibinfo {author} {\bibfnamefont {E.}~\bibnamefont
  {Uegaki}}, \bibinfo {author} {\bibfnamefont {Y.}~\bibnamefont {Abe}},
  \bibinfo {author} {\bibfnamefont {S.}~\bibnamefont {Okabe}}, \ and\ \bibinfo
  {author} {\bibfnamefont {H.}~\bibnamefont {Tanaka}},\ }\href {\doibase
  10.1143/PTP.62.1621} {\bibfield  {journal} {\bibinfo  {journal} {Progress of
  Theoretical Physics}\ }\textbf {\bibinfo {volume} {62}},\ \bibinfo {pages}
  {1621} (\bibinfo {year} {1979})}\BibitemShut {NoStop}%
\bibitem [{\citenamefont {Kamimura}(1981)}]{KAMIMURA1981456}%
  \BibitemOpen
  \bibfield  {author} {\bibinfo {author} {\bibfnamefont {M.}~\bibnamefont
  {Kamimura}},\ }\href {\doibase https://doi.org/10.1016/0375-9474(81)90182-2}
  {\bibfield  {journal} {\bibinfo  {journal} {Nuclear Physics A}\ }\textbf
  {\bibinfo {volume} {351}},\ \bibinfo {pages} {456 } (\bibinfo {year}
  {1981})}\BibitemShut {NoStop}%
\bibitem [{\citenamefont {Funaki}\ \emph {et~al.}(2003)\citenamefont {Funaki},
  \citenamefont {Tohsaki}, \citenamefont {Horiuchi}, \citenamefont {Schuck},\
  and\ \citenamefont {R\"opke}}]{PhysRevC.67.051306}%
  \BibitemOpen
  \bibfield  {author} {\bibinfo {author} {\bibfnamefont {Y.}~\bibnamefont
  {Funaki}}, \bibinfo {author} {\bibfnamefont {A.}~\bibnamefont {Tohsaki}},
  \bibinfo {author} {\bibfnamefont {H.}~\bibnamefont {Horiuchi}}, \bibinfo
  {author} {\bibfnamefont {P.}~\bibnamefont {Schuck}}, \ and\ \bibinfo {author}
  {\bibfnamefont {G.}~\bibnamefont {R\"opke}},\ }\href {\doibase
  10.1103/PhysRevC.67.051306} {\bibfield  {journal} {\bibinfo  {journal} {Phys.
  Rev. C}\ }\textbf {\bibinfo {volume} {67}},\ \bibinfo {pages} {051306}
  (\bibinfo {year} {2003})}\BibitemShut {NoStop}%
\bibitem [{\citenamefont {Suzuki}\ and\ \citenamefont
  {Ikeda}(1974)}]{10.1143/PTP.51.1621}%
  \BibitemOpen
  \bibfield  {author} {\bibinfo {author} {\bibfnamefont {Y.}~\bibnamefont
  {Suzuki}}\ and\ \bibinfo {author} {\bibfnamefont {K.}~\bibnamefont {Ikeda}},\
  }\href {\doibase 10.1143/PTP.51.1621} {\bibfield  {journal} {\bibinfo
  {journal} {Progress of Theoretical Physics}\ }\textbf {\bibinfo {volume}
  {51}},\ \bibinfo {pages} {1621} (\bibinfo {year} {1974})}\BibitemShut
  {NoStop}%
\bibitem [{\citenamefont {Suzuki}(1976{\natexlab{a}})}]{10.1143/PTP.55.1751}%
  \BibitemOpen
  \bibfield  {author} {\bibinfo {author} {\bibfnamefont {Y.}~\bibnamefont
  {Suzuki}},\ }\href {\doibase 10.1143/PTP.55.1751} {\bibfield  {journal}
  {\bibinfo  {journal} {Progress of Theoretical Physics}\ }\textbf {\bibinfo
  {volume} {55}},\ \bibinfo {pages} {1751} (\bibinfo {year}
  {1976}{\natexlab{a}})}\BibitemShut {NoStop}%
\bibitem [{\citenamefont {Suzuki}(1976{\natexlab{b}})}]{10.1143/PTP.56.111}%
  \BibitemOpen
  \bibfield  {author} {\bibinfo {author} {\bibfnamefont {Y.}~\bibnamefont
  {Suzuki}},\ }\href {\doibase 10.1143/PTP.56.111} {\bibfield  {journal}
  {\bibinfo  {journal} {Progress of Theoretical Physics}\ }\textbf {\bibinfo
  {volume} {56}},\ \bibinfo {pages} {111} (\bibinfo {year}
  {1976}{\natexlab{b}})}\BibitemShut {NoStop}%
\bibitem [{\citenamefont {Arima}\ \emph {et~al.}(1967)\citenamefont {Arima},
  \citenamefont {Horiuchi},\ and\ \citenamefont {Sebe}}]{ARIMA1967129}%
  \BibitemOpen
  \bibfield  {author} {\bibinfo {author} {\bibfnamefont {A.}~\bibnamefont
  {Arima}}, \bibinfo {author} {\bibfnamefont {H.}~\bibnamefont {Horiuchi}}, \
  and\ \bibinfo {author} {\bibfnamefont {T.}~\bibnamefont {Sebe}},\ }\href
  {\doibase https://doi.org/10.1016/0370-2693(67)90499-6} {\bibfield  {journal}
  {\bibinfo  {journal} {Physics Letters B}\ }\textbf {\bibinfo {volume} {24}},\
  \bibinfo {pages} {129 } (\bibinfo {year} {1967})}\BibitemShut {NoStop}%
\bibitem [{\citenamefont {Ikeda}\ \emph {et~al.}(1972)\citenamefont {Ikeda},
  \citenamefont {Marumori}, \citenamefont {Tamagaki},\ and\ \citenamefont
  {Tanaka}}]{doi:10.1143/PTPS.52.1}%
  \BibitemOpen
  \bibfield  {author} {\bibinfo {author} {\bibfnamefont {K.}~\bibnamefont
  {Ikeda}}, \bibinfo {author} {\bibfnamefont {T.}~\bibnamefont {Marumori}},
  \bibinfo {author} {\bibfnamefont {R.}~\bibnamefont {Tamagaki}}, \ and\
  \bibinfo {author} {\bibfnamefont {H.}~\bibnamefont {Tanaka}},\ }\href
  {\doibase 10.1143/PTPS.52.1} {\bibfield  {journal} {\bibinfo  {journal}
  {Progress of Theoretical Physics Supplement}\ }\textbf {\bibinfo {volume}
  {52}},\ \bibinfo {pages} {1} (\bibinfo {year} {1972})},\ \bibinfo {note} {and
  references therein}\BibitemShut {NoStop}%
\bibitem [{\citenamefont {Bisoi}\ \emph {et~al.}(2013)\citenamefont {Bisoi},
  \citenamefont {Sarkar}, \citenamefont {Sarkar}, \citenamefont {Ray},
  \citenamefont {Basu}, \citenamefont {Kanjilal}, \citenamefont {Nag},
  \citenamefont {Selvakumar}, \citenamefont {Goswami}, \citenamefont
  {Madhavan}, \citenamefont {Muralithar},\ and\ \citenamefont
  {Bhowmik}}]{PhysRevC.88.034303}%
  \BibitemOpen
  \bibfield  {author} {\bibinfo {author} {\bibfnamefont {A.}~\bibnamefont
  {Bisoi}}, \bibinfo {author} {\bibfnamefont {M.~S.}\ \bibnamefont {Sarkar}},
  \bibinfo {author} {\bibfnamefont {S.}~\bibnamefont {Sarkar}}, \bibinfo
  {author} {\bibfnamefont {S.}~\bibnamefont {Ray}}, \bibinfo {author}
  {\bibfnamefont {M.~R.}\ \bibnamefont {Basu}}, \bibinfo {author}
  {\bibfnamefont {D.}~\bibnamefont {Kanjilal}}, \bibinfo {author}
  {\bibfnamefont {S.}~\bibnamefont {Nag}}, \bibinfo {author} {\bibfnamefont
  {K.}~\bibnamefont {Selvakumar}}, \bibinfo {author} {\bibfnamefont
  {A.}~\bibnamefont {Goswami}}, \bibinfo {author} {\bibfnamefont
  {N.}~\bibnamefont {Madhavan}}, \bibinfo {author} {\bibfnamefont
  {S.}~\bibnamefont {Muralithar}}, \ and\ \bibinfo {author} {\bibfnamefont
  {R.~K.}\ \bibnamefont {Bhowmik}},\ }\href {\doibase
  10.1103/PhysRevC.88.034303} {\bibfield  {journal} {\bibinfo  {journal} {Phys.
  Rev. C}\ }\textbf {\bibinfo {volume} {88}},\ \bibinfo {pages} {034303}
  (\bibinfo {year} {2013})}\BibitemShut {NoStop}%
\bibitem [{\citenamefont {Inakura}\ \emph {et~al.}(2002)\citenamefont
  {Inakura}, \citenamefont {Mizutori}, \citenamefont {Yamagami},\ and\
  \citenamefont {Matsuyanagi}}]{Inakura2002261}%
  \BibitemOpen
  \bibfield  {author} {\bibinfo {author} {\bibfnamefont {T.}~\bibnamefont
  {Inakura}}, \bibinfo {author} {\bibfnamefont {S.}~\bibnamefont {Mizutori}},
  \bibinfo {author} {\bibfnamefont {M.}~\bibnamefont {Yamagami}}, \ and\
  \bibinfo {author} {\bibfnamefont {K.}~\bibnamefont {Matsuyanagi}},\ }\href
  {\doibase DOI: 10.1016/S0375-9474(02)01164-8} {\bibfield  {journal} {\bibinfo
   {journal} {Nuclear Physics A}\ }\textbf {\bibinfo {volume} {710}},\ \bibinfo
  {pages} {261 } (\bibinfo {year} {2002})}\BibitemShut {NoStop}%
\bibitem [{\citenamefont {Bender}\ \emph {et~al.}(2003)\citenamefont {Bender},
  \citenamefont {Flocard},\ and\ \citenamefont {Heenen}}]{PhysRevC.68.044321}%
  \BibitemOpen
  \bibfield  {author} {\bibinfo {author} {\bibfnamefont {M.}~\bibnamefont
  {Bender}}, \bibinfo {author} {\bibfnamefont {H.}~\bibnamefont {Flocard}}, \
  and\ \bibinfo {author} {\bibfnamefont {P.~H.}\ \bibnamefont {Heenen}},\
  }\href {\doibase 10.1103/PhysRevC.68.044321} {\bibfield  {journal} {\bibinfo
  {journal} {Phys. Rev. C}\ }\textbf {\bibinfo {volume} {68}},\ \bibinfo
  {pages} {044321} (\bibinfo {year} {2003})}\BibitemShut {NoStop}%
\bibitem [{\citenamefont {Kanada-En'yo}\ and\ \citenamefont
  {Horiuchi}(1995)}]{PTP.93.115}%
  \BibitemOpen
  \bibfield  {author} {\bibinfo {author} {\bibfnamefont {Y.}~\bibnamefont
  {Kanada-En'yo}}\ and\ \bibinfo {author} {\bibfnamefont {H.}~\bibnamefont
  {Horiuchi}},\ }\href {\doibase 10.1143/PTP.93.115} {\bibfield  {journal}
  {\bibinfo  {journal} {Progress of Theoretical Physics}\ }\textbf {\bibinfo
  {volume} {93}},\ \bibinfo {pages} {115} (\bibinfo {year} {1995})}\BibitemShut
  {NoStop}%
\bibitem [{\citenamefont {Kimura}(2004)}]{PhysRevC.69.044319}%
  \BibitemOpen
  \bibfield  {author} {\bibinfo {author} {\bibfnamefont {M.}~\bibnamefont
  {Kimura}},\ }\href {\doibase 10.1103/PhysRevC.69.044319} {\bibfield
  {journal} {\bibinfo  {journal} {Phys. Rev. C}\ }\textbf {\bibinfo {volume}
  {69}},\ \bibinfo {pages} {044319} (\bibinfo {year} {2004})}\BibitemShut
  {NoStop}%
\bibitem [{\citenamefont {Taniguchi}\ \emph {et~al.}(2004)\citenamefont
  {Taniguchi}, \citenamefont {Kimura},\ and\ \citenamefont
  {Horiuchi}}]{PTP.112.475}%
  \BibitemOpen
  \bibfield  {author} {\bibinfo {author} {\bibfnamefont {Y.}~\bibnamefont
  {Taniguchi}}, \bibinfo {author} {\bibfnamefont {M.}~\bibnamefont {Kimura}}, \
  and\ \bibinfo {author} {\bibfnamefont {H.}~\bibnamefont {Horiuchi}},\ }\href
  {\doibase 10.1143/PTP.112.475} {\bibfield  {journal} {\bibinfo  {journal}
  {Progress of Theoretical Physics}\ }\textbf {\bibinfo {volume} {112}},\
  \bibinfo {pages} {475} (\bibinfo {year} {2004})}\BibitemShut {NoStop}%
\bibitem [{\citenamefont {Taniguchi}(2016)}]{Taniguchi01102016}%
  \BibitemOpen
  \bibfield  {author} {\bibinfo {author} {\bibfnamefont {Y.}~\bibnamefont
  {Taniguchi}},\ }\href {\doibase 10.1093/ptep/ptw140} {\bibfield  {journal}
  {\bibinfo  {journal} {Progress of Theoretical and Experimental Physics}\
  }\textbf {\bibinfo {volume} {2016}},\ \bibinfo {pages} {103D01} (\bibinfo
  {year} {2016})}\BibitemShut {NoStop}%
\bibitem [{\citenamefont {Chen}\ \emph {et~al.}(2011)\citenamefont {Chen},
  \citenamefont {Cameron},\ and\ \citenamefont {Singh}}]{CHEN20112715}%
  \BibitemOpen
  \bibfield  {author} {\bibinfo {author} {\bibfnamefont {J.}~\bibnamefont
  {Chen}}, \bibinfo {author} {\bibfnamefont {J.}~\bibnamefont {Cameron}}, \
  and\ \bibinfo {author} {\bibfnamefont {B.}~\bibnamefont {Singh}},\ }\href
  {\doibase https://doi.org/10.1016/j.nds.2011.10.001} {\bibfield  {journal}
  {\bibinfo  {journal} {Nuclear Data Sheets}\ }\textbf {\bibinfo {volume}
  {112}},\ \bibinfo {pages} {2715 } (\bibinfo {year} {2011})}\BibitemShut
  {NoStop}%
\end{thebibliography}%

\end{document}